\documentclass{article}

\usepackage{arxiv}

\usepackage[utf8]{inputenc} 
\usepackage[T1]{fontenc}    
\usepackage{hyperref}       
\usepackage{url}            
\usepackage{booktabs}       
\usepackage{amsfonts}       
\usepackage{nicefrac}       
\usepackage{microtype}      
\usepackage{lipsum}		
\usepackage{graphicx}
\usepackage{natbib}
\usepackage{doi}
\usepackage{chemformula}   

\title{PEPICO analysis of catalytic reactor effluents towards quantitative isomer discrimination: DME conversion over a ZSM-5 zeolite}

\author{ 
   {
   \hspace{1mm}Morsal~Babayan}$^*$\\
	Nano and Molecular Systems Research Unit\\
	University of Oulu\\
	Oulu, Finland \\
	\texttt{morsal.babayan@oulu.fi} \\
	\And
    {
    \hspace{1mm}Evgeniy~Redekop} \\
	Department of Chemistry\\
    Centre for Materials Science and Nanotechnology (SMN)\\
	University of Oslo\\
	Oslo, Norway \\
	\texttt{evgeniy.redekop@smn.uio.no} \\
	\And
    {
    \hspace{1mm}Esko~Kokkonen} \\
	MAX IV Laboratory\\
	Lund University\\
	Lund, Sweden\\
	\And
    {
    \hspace{1mm}Unni~Olsbye} \\
	Department of Chemistry\\
    Centre for Materials Science and Nanotechnology (SMN)\\
	University of Oslo\\
	Oslo, Norway \\
	\And
   {
   \hspace{1mm}Marko~Huttula}\\
	Nano and Molecular Systems Research Unit\\
	University of Oulu\\
	Oulu, Finland \\
	\And
   {
   \hspace{1mm}Samuli~Urpelainen}$^*$\\
	Nano and Molecular Systems Research Unit\\
	University of Oulu\\
	Oulu, Finland \\
	\texttt{samuli.urpelainen@oulu.fi} \\
}

\date{}

\renewcommand{\headeright}{}
\renewcommand{\shorttitle}{PEPICO analysis of catalytic reactor effluents at MAX IV}

\hypersetup{
pdftitle={A template for the arxiv style},
pdfsubject={q-bio.NC, q-bio.QM},
pdfauthor={Morsal~Babayan, Evgeniy~Redekop, Esko Kokkonen, Unni Olsbye, Marko Huttula, Samuli Urpelainen},
pdfkeywords={First keyword, Second keyword, More},
}

\begin{document}
\maketitle

\begin{abstract}
The Methanol-To-Hydrocarbons (MTH) process involves the conversion of methanol, a C1 feedstock that can be produced from green sources, into hydrocarbons using shape-selective microporous acidic catalysts - zeolite and zeotypes \cite{olsbye2012}. 
This reaction yields a complex mixture of species, some of which are highly reactive and/or present in several isomeric forms, posing significant challenges for effluent analysis. Conventional gas-phase chromatography (GC) is typically employed for the analysis of reaction products in laboratory flow reactors. However, GC is not suitable for the detection of highly reactive intermediates such as ketene or formaldehyde and is not suitable for kinetic studies under well-defined low pressure conditions. Photoelectron–photoion coincidence (PEPICO) spectroscopy has emerged as a powerful analytical tool for unraveling complex compositions of catalytic effluents \cite{hemberger2020new}, but its availability is limited to a handful of facilities worldwide. Herein, PEPICO analysis of catalytic reactor effluents has been implemented at the FinEstBeAMS beamline of MAX IV Laboratory. 
The conversion of dimethyl ether (DME) on a zeolite catalyst (ZSM-5-MFI27) is used as a prototypical model reaction producing a wide distribution of hydrocarbon products. 
Since in zeolites methanol is quickly equilibrated with DME, this reaction can be used to probe vast sub-networks of the full MTH process, while eliminating or at least slowing down methanol-induced secondary reactions and catalyst deactivation. 
Quantitative discrimination of xylene isomers in the effluent stream is achieved by deconvoluting the coincidence photoelectron spectra.
\end{abstract}


\section{Introduction}

Catalytic transformations of hydrocarbons and other organic molecules inside microporous acidic zeolites drive many large scale industrial chemical processes with enormous combined economic and environmental impact \cite{cejka2010}. 
Some of zeolite-mediated catalytic processes are also key candidates for implementing more sustainable transformations of diversified raw materials and renewable energy into vital energy carriers and chemical intermediates {\cite{li2021}.
Highly convoluted reaction networks and microporous transport phenomena involved in catalysis by zeolites often present formidable experimental challenges. 
Complex effluent streams produced by zeolite catalysts may contain dozens of compounds, many of them present in several isomeric forms. Gas chromatography (GC) is a commonly used technique for effluent analysis in laboratory experiments that are typically conducted at ambient or above-ambient pressure conditions. 
However, GC is limited to the analysis of gas samples extracted from ambient or near ambient pressure conditions and is not suitable for capturing highly reactive closed and open shell gaseous intermediates, e.g. formaldehyde, ketenes, or radicals.
Kinetic measurements under well-defined, low pressure reaction conditions have emerged as an important source of mechanistic information because these highly reactive intermediates can be more readily detected and quantified by mass-spectrometry 
\cite{brogaard2014, batchu2017, omojola2021}. 
Likewise, low pressure operation prolongs the catalyst lifetime by minimizing secondary reactions and coking, thus expanding the range of catalyst states that are amenable for precise kinetic characterization \cite{redekop2020temporal}.
Isomer-selective analytics at low-pressure conditions would open new avenues for advanced kinetic characterization of catalytic reactions at model operating conditions that facilitate in-depth mechanistic studies.

Photoionization Mass-Spectrometry (PIMS) provides a suitable analytical platform for advanced mechanistic investigations, whereby an analyte is ionized by an incident photon and the resulting photoions are detected. 
PIMS achieves it’s full analytical potential when energy of the incident light can be varied in the Vacuum Ultraviolet (VUV) range (6-42 eV), which typically requires a synchrotron radiation source. 
In comparison with conventional electron ionization MS and gas chromatography, PIMS-based effluent analysis offers unique analytical advantages including (i) better sensitivity and resolving power even for highly reactive species, (ii) applicability across a broad range of operating conditions, from vacuum to ambient pressure (with differential pumping), and (iii) in many cases, when performed with high photon energy resolution, can offer isomer selectivity, which is particularly valuable for organic reactions. 
However, in order to have a better insight into the gaseous composition, molecular photofragmentation must be considered in considerable depth \cite{kooser2020gas}. Although the electron and ion spectroscopy methods alone play a major part in studying molecular photofragmentation, in particular, recording and analyzing electrons and ions originating from the same photoionization event can lead to more complete understanding of the photofragmentation process and the composition of the effluent gas stream. Taking the type of the detected particles into account, coincidence techniques may be classified as photoelectron–photoion coincidence (PEPICO), photoion–photoion coincidence (PIPICO), photoelectron–photoion–photoion coincidence (PEPIPICO), photoelectron–photoelectron coincidence and photoion–neutral coincidence \cite{arion2015coincidence}. PEPICO and Threshold PIMS have been most instrumental in unraveling the reaction mechanisms of Methanol-To-Hydrocarbons (MTH), catalytic pyrolysis, and oxychlorination processes. With the aid of \textit{in situ} synchrotron radiation PIMS, Wen et al. \cite{wen2020} have detected formaldehyde (\ch{HCHO}), an active intermediate, during the MTH reaction over two various catalysts.
Recently, Cesarini et al. \cite{cesarini2022} utilized \textit{operando} PEPICO spectroscopy to investigate reaction pathways for MTH and MCTH (methyl chloride-to-hydrocarbons) over the H-ZSM-5 zeolite catalyst. 
By using this technique, they directly observed short-lived active intermediates, such as ketene and methyl radicals. Despite the increasing interest in applying these methods to investigate catalytic reactions, few experimental setups exist which combine catalytic microreactors with PIMS-based effluent analysis. 

Herein, we report the establishment of such instrument at the gas-phase endstation of the FinEstBeAMS beamline of MAX IV Laboratory. 
Currently there are only a handful of synchrotron beamlines providing capabilities for the PIMS-based analysis of catalytic reactor effluents, including FinEstBeAMS. 
Features of some of them are are compared in Table \ref{tab:facilities}. 
The main differences between these setups are the accessible energy ranges and the types of available detectors. 
The VUV beamline at SLS offers photons in the 3–150 eV range and is capable of both photoion and photoelectron analysis \cite{SLS}.
The BL03U beamline at NSRL has a relatively narrow photon energy range of 5–21 eV and only detects ions \cite{NSRL2016}.
The DESIRS beamline at SOLEIL covering the VUV range (5-40 eV) is equipped with a double imaging photoelectron photoion coincidence (i\textsuperscript{2}PEPICO) spectrometer with two position sensitive detector to detect both photoelectrons and photoions \cite{tang2015}.
In comparison, FinEstBeMS covers a broad range of photon energies 4.4–1000 eV, which enables both valence and core ionization, and features  ion and electron analysis as well as coincidence experiments.

The conversion of Methanol and/or dimethyl ether (DME) to hydrocarbons, abbreviated MTH and DTH, respectively, on acidic zeolites and zeotypes offers a prototypical example of a catalytic reaction which produces compositionally and isomerically complex effluents. MTH/DTH are promising industrial routes towards hydrocarbon fuels and platform chemicals, which can accommodate diverse feedstocks including bio-gas and captured \ch{CO2} \cite{xie2023}. 
In acidic zeolite or zeotype catalysts, an equilibrated mixture of DME and methanol react on Br{\o}nsted Acid Sites (BAS) to produce larger hydrocarbon molecules including alkenes, alkanes, and aromatics. 
The reaction proceeds through the dual-cycle Hydrocarbon Pool (HCP) mechanism in which several pathways can be distinguished. 
The first C-C bonds are formed from DME and/or methanol via Surface Methoxy Species (SMS) and highly reactive intermediates such as formaldehyde or ketene, eventually leading to C2-C3 alkenes. 
These primary products are then repeatedly methylated by SMS (stepwise pathway) or gaseous DME/methanol to form C3-C5+ alkenes – the so-called alkene cycle. 
The product distribution is further controlled by co-occuring cracking reactions. Hydrogen transfer reactions between methanol and alkenes, also catalyzed by BAS, lead to the formation of alkanes and aromatics. 
The latter can sustain an independent cycle of sequential methylation and cracking – the aromatics cycle. 
At steady-state, the population of \textit{in situ} generated alkene and aromatics intermediates termed the HCP resides within the catalyst and mediates continuous catalytic production of products from DME/methanol, in parallel with the hydrogen transfer and isomerisation reactions. 
Moreover, methyl radicals were also detected in the reaction medium, suggesting that the underlying chemistry may be even more complex than previously thought. 
Eventually, the growth of large polyaromatic molecules and coke occludes the microporous space and deactivates the catalyst. 

Catalytic performance, i.e., activity, selectivity, and stability, is ultimately controlled by a multitude of factors related to the catalyst structure and operating conditions. 
In order to establish the structure-performance relationships and optimize the catalytic materials and reactions, it is imperative to better understand the kinetics of different reaction pathways and how they are affected by variations of the materials composition and structure. 
However, disentangling the distinct reaction pathways and individual reaction steps is a challenging experimental task, given the aforementioned complexity. 
PIMS-based methodologies for the reaction analysis have already provided essential mechanistic insights into MTO chemistry, and this provides the main motivation for the current study - to showcase the new \textit{operando} PEPICO capabilities at the FinEstBeAMS beamline in the context of this important reaction. 
In particular, we demonstrate PEPICO analysis of DME conversion on a ZSM-5 catalyst with a particular emphasis on quantitative isomer discrimination for the product xylenes.

\section{Experiments}
In this study, the products of the DME reaction over the ZSM-5 zeolite catalyst at temperature of 375 \textdegree C are investigated in real-time, during the catalytic process, through \textit{operando} PEPICO spectroscopy. 
This investigation has been performed using coincidence setup at the gas-phase endstation (GPES) of the FinEstBeAMS beamline at the MAX IV Laboratory (Lund, Sweden). 
Pure DME was fed into the reactor at a rate of $0.045$ sccm via a precision leak valve opening directly into the packed bed.
At this feed rate, the pressure in the analysis chamber remained at $5 \times 10^{-7}$ mbar. 
In this experiment, photons with an energy of $40$ eV were utilized as the ionization source.
The electron analyzer operates at pass energy of $100$ eV with the kinetic energy window centered at $28$ eV.
Hence, electrons with kinetic energies within the range of $23-33$ eV (equivalent to binding energies of $7-17$ eV) were captured.
Moreover, the measurement was conducted at the magic angle of 55\textdegree.
While the beamline has a sub-meV resolution at 40 eV photon energy \cite{parna2017}, the electron spectrometer resolution was approximately $600$ meV.
The electron count rate and random trigger frequency were kept at about $25$ Hz to have a better coincidence purity.
In total, over a duration of approximately 15 hours, nearly $2 \times 10^{6}$ triggers were detected, in which roughly $21 \%$ of them were coincidence triggers.
Coincidence-specific data handling was done with custom Igor Pro macros \cite{kukk2007}.

\subsection{Gas-phase endstation}
The FinEstBeAMS beamline is in the $1.5$ GeV storage ring providing photon energy in the range from ultraviolet to soft X-ray (i.e., $4.5$ to $1300$ eV) as well as variable polarization of synchrotron radiation.
Detailed information on the design and optical concept of the FinEstBeAMS beamline is given elsewhere \cite{parna2017, chernenko2021}. 
The FinEstBeAMS beamline consists of two separate branch lines. 
The GPES and the photoluminescence endstation (PLES) are in the same branch, while the solid-state endstation (SSES) is located in the other one.

The GPES is developed for electron and ion spectroscopy as well as photoelectron–photoion coincidence spectroscopy of low-density matter.
Kooser et al.(2020) describes the GPES in more details. 
Briefly, in this apparatus a modified hemispherical electron energy analyser (SCIENTA R4000) equipped with a fast resistive anode position-sensitive detector is utilized to detect photoelectrons. 
Moreover, a Wiley-McLaren ion time of flight (TOF) mass spectrometer with delay-line position-sensitive detector detects ions produced throughout the ionization.
In order to calibrate the TOF spectrum to masses, the following formula is utilized $m/z = (TOF-T_{0})^{2}/C^{2}$. 
Here, the calibration parameters are determined by using the peaks corresponding to masses $2$ (\ch{H2}) and $91$ (\ch{C7H7}) as the initial guess. 
Using these initial values, we calculate $T_{0} = 1715$ ns and $C = -11.59$ ns(e/u)$^{1/2}$ which ensures that all other TOF peaks fall into the correct masses.

\subsection{Reactor}
A portable, flange-mounted catalytic packed-bed reactor (length $50$ mm, internal diameter $4$ mm) was interfaced directly with the GPES end-station, as shown in Fig. \ref{fig:GPES}. 
In brief, 20 mg of a catalytic sample was packed in the middle of the reactor as a thin ($2$ mm) layer sandwiched between two inert zones that were packed with quartz particles of the same sieve fraction. 
The reactor was resistively heated, while the temperature was monitored by a K-type thermocouple positioned in the middle of the catalytic bed. 
Gaseous reactants were fed into the reactor through a calibrated leak valve, and the effluent was allowed to freely enter the analysis chamber where the gas was pumped away by three turbo molecular pumps with total capacity of about $1300$ l/s.
The schematic of the reactor connected to the GPES is presented in Fig. \ref{fig:GPES}.

\subsection{Materials and reagents}
A commercial ZSM-5-MFI-27 zeolite was purchased from Sud Chemie. 
To obtain the acidic form of the catalyst, the as-received material was ion-exchanged with \ch{NH4NO3}, extensively washed, and calcined in static air at 550 \textdegree C for 10 hours. 
Then, the catalyst was pressed into pellets that were sieved to $250 < d_{p} < 400$ $\mu$m size fraction, which were subsequently packed into the reactor. 
Before the experiment, the catalyst was maintained in vacuum at 550 \textdegree C for 30 minutes to desorb the residual water. 
The detailed procedure for the catalyst preparation and extensive standard characterization data can be found in \cite{rojo2018}.
Basic physico-chemical properties were determined to be as follows: Si/Al ratio of $15$, BAS (Br{\o}nsted acid sites) concentration of 0.87 mmol/g, crystal size of 2-6 $\mu$m, and BET (Brunauer-Emmett-Teller) surface area of 398 m$^2$/g.

\subsection{PEPICO}
In this study, \textit{operando} PEPICO spectroscopy is used which aside from being isomer-selective, is able to qualitatively differentiate short- and long-lived species as well.
In PEPICO spectroscopy, both the photoelectron and the photoion generated via the ionization are detected. 
For regular PEPICO spectroscopy, the kinetic energy of electrons in conjunction with cations with same mass per charge ratio provides the mass-resolved photoelectron spectrum (ms-PES) for that specific cation.
However, in threshold-PEPICO spectroscopy, photoions in coincidence with electrons having near-zero kinetic energy are collected.
Therefore, mass-selected threshold photoelectron spectrum (ms-TPES) is provided for a particular mass per charge ratio by threshold-PEPICO \cite{bodi2013}.
To detect different isomers using threshold-PEPICO, tunable light sources with sufficiently high resolution are required.
Hence, the regular PEPICO technique (from hereon PEPICO), readily available at FinEstBeAMS, is considered in this study.

The products of the reaction and the remaining reactant (when the conversion rate is less than 100\%) leaving the microreactor are ionized by the photon beam.
It leads to the generation of ions and electrons in the extraction region of the time-of-flight(TOF) mass spectrometer. 
When an electron is detected by the electron analyzer, a signal is generated to initiate a pulsed electric field in the extraction region of the mass spectrometer. 
Since the electron mass is negligible comparing to the ions, the flight times of electrons are significantly lower than the ions, therefore it is reasonable to assume that the formation of the ions and the detection of the electron occur concomitantly. 
Eventually, the electron-ion pairs associated with the same photoionization event can be distinguished by correlating the detected electrons and ions. 

\subsubsection{True/false coincidences}
Although one electron is detected each time, ionizing more than one atom or molecule throughout the ionization process is possible which leads to detect the electron/ion pairs that are not coming from the same event.
Such events are known as false coincidences.
By applying low ionization rate, the probability of detecting the electron/ion pairs arising from the same event (true coincidences) increases, as the generated electron/ion pairs will be well-separated in time \cite{bodi2013}.
Furthermore, less than $100$ percentage detection efficiency of the electron and ion detectors (which is lower for the electron) contributes to the false coincidences. 
To distinguish the ions generated by true coincidences,  subtracting the random ions coming from the false coincidences is required. 
Consequently, a reference random coincidence must be measured under exactly the same conditions.
Hence, an external pulse generator is used to create artificial random triggers besides electron triggers.
For those random triggers, all measured coincidences would be false coincidences. 
Eventually, subtracting the random coincidences from total coincidences coming from electron triggers, leads to true coincidences. 
Previously \cite{prumper2007}, coincidence experiments and the random coincidences subtraction method have been described in more details. 
Here, random triggers generated by an external pulse generator with $25$ Hz frequency is utilized. 
The TOF spectra from electron-triggered, random, and true coincidences are depicted in Fig. 1S.

\subsubsection{Electron spectra}
PEPICO measurements provide both the TOF of ions, revealing the mass per charge ratio of the ions, as well as the kinetic energy of the ejected electrons.
This technique is commonly used to analyze the unimolecular dissociation.
However, analyzing catalytic reactions often involves dealing with multiple products and leftover reactants, making it difficult to distinguish parent molecules from ionization fragments.
In these cases, coincidence ion yield photoelectron spectra (CIY-PES), analogous to ms-PES referred by \cite{cesarini2022}, can be used to provide additional information.
For atoms, photoelectron spectroscopy shows the binding energies of the electrons, whereas for molecules, it reveals vibrational and rotational excitations as well.
Photoelectron spectroscopy reveals key features of the original molecular orbitals from which the electrons are emitted.  
On the other hand, isomers of a molecule have distinct electronic structures due to differences in the relative positioning of substituent groups, such as methyl groups in xylene, which can alter electronic density and molecular symmetry.
This makes photoelectron spectroscopy a useful tool not only for distinguishing between different molecules but also for identifying various isomers of a molecule.
For example, the photoelectron spectra obtained for m- and p-xylene by \cite{koenig1974} using He(I) radiation revealed that the first ionic states of these isomers have distinct vertical ionization potentials that can be utilized for differentiation purposes. 
Likewise, unique features observed in the photoelectron spectra of cis-, trans-, and iso-butene can serve as distinctive markers to differentiate between these isomers \cite{ying1993}.
Therefore, utilizing CIY-PES can be a valuable tool for differentiating between ions with identical mass/fragments.

\section{Results and discussion}
\subsection{TOF spectra}\label{subsec;TOF}
The detected ions, representing the effluent stream during the conversion of dimethyl ether (DME) on the ZSM-5 catalyst, are generated mainly by single valence ionisation, as the cross section for direct double photoionization is low.
Therefore, for the sake of abbreviation, instead of mass per charge ratio, only mass is going to be used.

A broad range of ions, covering masses from $1$ to $156$, are detected in the effluent stream.
Table \ref{table:ions} lists the masses detected in true coincidences and the possible corresponding molecular ions.
The calibrated TOF spectrum of true ions detected in coincidence with electrons in the binding energy range of ca. $7$ to $17$ eV is depicted in Fig. \ref{fig:map}b. 
Due to the negligible intensity observed for ions with masses greater than $120$, they are excluded from the figure.
Furthermore, it is noteworthy to highlight that the comparison between the TOF spectrum for DME flowing over the blank reactor closely resembles the reference spectrum provided in NIST Chemistry WebBook \cite{NIST}.
This correspondence effectively mitigates concerns regarding the potential reactivity of DME under these experimental conditions.
The TOF spectra of a mixture of Ar and DME (3:1 ratio) flowing over the empty reactor is plotted in Fig. 2S for two different rector temperatures.

In the current experiment, DME is not present in the effluent and its complete conversion is attained under the considered conditions.
Feng et al. \cite{feng2001} reported that at $40$ eV photon energy for DME, the branching ratios of ions with masses $46$ and $45$ were about $13.6\%$ and $22.5\%$, respectively. 
However, ions with these masses are not detected in this study (as shown in Table \ref{table:ions}). 

In Fig. \ref{fig:map}b, the peak with the highest intensity corresponds to \ch{H2O+} (mass $18$) - an expected result, considering that water is a major product of DME conversion on zeolite catalysts \cite{Chang1983}. The peak with the second highest intensity corresponds to the mass $28$, which can be attributed to \ch{N2+}, \ch{CO+}, and/or \ch{C2H4+} cations. 
However, it is not possible to determine from the TOF spectrum alone whether all of these cations contribute to this peak or not.
Moreover, the peaks for masses $16$ and $15$ exhibit noticeable intensities in the recorded spectrum. 
\ch{CH3+} cation is the only plausible candidate for the peak with mass $15$, given that the effluent is expected to contain hydrocarbons, water, and residual oxygenates (DME and methanol). However, it is possible that both \ch{CH4+} and \ch{O+} contribute to the peak observed at the mass $16$.
Based on the peaks observed at masses $18$, $28$, $32$, and $44$, it is possible that the \ch{O+} cation detected in the experiment originates from the fragmentation of \ch{H2O}, \ch{CO}, \ch{O2}, \ch{CH3OH}, or \ch{CO2} molecules, respectively.
Nonetheless, the absence of a peak at mass $31$ implies that \ch{CH3OH} is not present in the effluent stream, which is consistent with a complete conversion of oxygenates on ZSM-5 catalyst at these conditions.
For the peak at mass $44$, both \ch{CO2+}, and \ch{C3H8+} are the possible cations that we can not discriminate only by TOF measurement. 
There is a group of small peaks at masses $39-42$, which originate possibly from photofragmentation of propylene.

The TOF spectra show a pair of peaks at masses $91$ and $92$, which correspond to \ch{C7H7+} and \ch{C7H8+}, respectively, and are the major fragments of toluene.
Moreover, apart from \ch{C8H9+} and \ch{C8H10+}, \ch{C7H7+} is a main fragment of xylene.
The pair of peaks observed at $105$ and $106$ represent  \ch{C8H9+} and \ch{C8H10+} cations, respectively.
Although the TOF spectrum provides valuable information about the potential parent molecules in the effluent stream, toluene and xylene in this case, it is not possible to determine the specific origin of the \ch{C7H7+} cation.

In general, conventional mass spectroscopy techniques, including TOF, have limitations when it comes to analyzing the complex mixtures or molecules with identical masses
(like isomers) or those exhibiting similar fragments.

Hence, alternative techniques like PEPICO are employed to overcome these limitations and provide more detailed information about the parent molecules.
This is achieved by recording the kinetic energy of the detected electrons and extracting CIY-PES for each detected masses. 
These spectra can then be compared with the previously reported spectra of potential parent molecules to validate the identification achieved by mass spectrometry.

\subsection{PEPICO}
Fig. \ref{fig:map} shows an electron-energy-resolved PEPICO map of electronic states with binding energies between 8-17 eV. 
This figure shows a map of the event intensities for all ion TOF and electron energy pairs.
The horizontal axis corresponds to the electron hit position energies (which is calibrated to the electron binding energy in the top panel) and the vertical axis corresponds to the simultaneously detected ion flight times. 
This map provides an overview of the fragmentation patterns 
by associating specific electron binding energies with their corresponding positively charged fragments.
False coincidences have been removed and slight smoothing is applied for better visualization. 
While the general trend can be seen in Fig. \ref{fig:map}, a more detailed analysis should be conducted using the CIY-PES, which is given in the next section.

\subsection{CIY photoelectron spectra} \label{subsec;CIY}
As described in section \ref{subsec;TOF}, the TOF spectrum furnishes valuable insights into the mass distribution of ions within the chamber, along with their relative abundances, aiding in compound identification to a certain extent.
However, when it comes to molecules with identical masses or a mixture of samples exhibiting similar ionization fragmentation patterns, TOF is limited. 
For instance, the mass $91$ ionization fragment is predominant in both xylene and toluene, making them indistinguishable in a mixture using only TOF.
CIY-PES within the framework of PEPICO addresses such challenges by providing additional information about the emitted photoelectrons in coincidence with the mass of interest.

Fig. \ref{fig:map}a illustrates the photoelectron spectrum (PES) of all detected electrons in coincidence with true ions.
In this section, we aim to discriminate between the different species in the effluent stream by comparing the extracted CIY-PES with those previously reported for potential source compounds. 
To perform this comparative analysis, reference spectra are digitized using the WebPlotDigitizer online software \cite{Rohatgi2022}.

The TOF spectrum reveals two intense peaks at masses $15$ and $16$ (Fig. \ref{fig:map}b).
The peak at mass $15$ corresponds only to the \ch{CH3^+} cation, while that at mass $16$ can be attributed to \ch{CH4^+} and \ch{O+} cations. 
The potential parent molecules that contain oxygen and can give rise to \ch{O+} ion in this study are \ch{O2}, \ch{CO}, \ch{CO2}, water, and DME. 
However, the mass spectra of these molecules, as reported by the NIST Chemistry WebBook \cite{NIST}, would exhibit a considerably smaller peak at mass 16. 
In our dataset, DME undergoes complete conversion, and the molecular ions of \ch{O2}, \ch{CO}, and \ch{CO2} have even lower intensities than mass $16$. 
Additionally, according to the NIST Chemistry WebBook \cite{NIST}, the mass 16 fragment of water is negligible.
Therefore, it is assumed that \ch{CH4+} is the main cation contributing to the peak in question.

Fig. \ref{fig:CIY-PES}a shows a comparison between the CIY-PES for masses $15$ and $16$, and their sum with the reference PES of \ch{CH4}. 
The CIY-PES of mass $16$ only covers a portion of the reference spectrum, while the CIY-PES of mass $15$ covers the remaining section. 
The sum of these two spectra results in a spectrum that closely resembles the reference spectrum of \ch{CH4} as reported by Kimura \cite{kimura1981}.
The ionization of a molecule can occur from different sites, including the bonding and antibonding orbitals. 
For \ch{CH4}, the main fragmentation channels are \cite{chang2017}:
\begin{equation}
    \ch{CH4} + h\nu \rightarrow  \ch{CH4+} + \ch{e-}
\end{equation}
\begin{equation}
    \ch{CH4} + h\nu \rightarrow  \ch{CH4+} + \ch{e-} \rightarrow  \ch{CH3+} + \ch{H} + \ch{e-}
\end{equation}
Therefore, aggregating the CIY-PESs collected with PEPICO for main fragments (in this case, \ch{CH3+} and \ch{CH4+}) can result in a spectrum that is more representative of the reference spectra (i.e., \ch{CH4} PES). 
Because, in
reference spectra, electrons from all ionization channels, including all originating from
both inner valence ionization and outermost shell ionization, are considered. Whereas,
in CIY-PES only electrons generated in a specific ionization channel are taken into account.

Mass $18$ exhibits the most prominent peak in the TOF spectrum, with water being the most likely species of origin, as stated previously.
To verify this, the extracted CIY-PES for mass $18$ is compared with the He(I) photoelectron spectrum of water in Fig. \ref{fig:CIY-PES}b \cite{kimura1981}. 
Despite the differences in the photon energies and the resolution, this figure demonstrates a favorable agreement between the measured and the reference spectra.

Based on the nature of the experiment, the most likely cations with mass $28$ are \ch{C2H4+} (a product of the reaction), \ch{CO+} (a chamber contaminant), and \ch{N2+} (an air leak). 
Comparison of the corresponding CIY-PES for mass $28$ with the reference He(I) photoelectron spectra of \ch{N2}, \ch{CO}, and \ch{C2H2} \cite{kimura1981} (see Fig. S4) reveals that \ch{N2} and \ch{CO} are present in the effluent stream. 
However, the reference spectrum for \ch{C2H4} displays a broad peak at binding energy of $14.66$ eV which is not captured by only the CIY-PES for mass $28$. 
According to previous studies \cite{NIST, stockbauer1975}, in addition to the molecular ion \ch{C2H4+}, pure \ch{C2H4} also fragments into \ch{C2H3+} and \ch{C2H2+} with masses $27$ and $26$, respectively. The combined CIY-PES for masses $26$, $27$, and $28$, indeed, demonstrate a better agreement with the reference photoelectron spectrum of pure \ch{C2H4}
because it includes all major fragmentation channels (as presented in Fig. \ref{fig:CIY-PES}c). 

The detection of \ch{N2} indicates the presence of air leakage into the chamber, which is also confirmed by the observation of \ch{O2} with a distinct peak at mass $32$ in the TOF spectrum (Fig. \ref{fig:map}b) and the corresponding comparison of CIY-PES with the reference photoelectron spectrum for \ch{O2} \cite{kimura1981} (see Fig. S5).

Ions with masses of $41$ and $42$ are also detected, albeit with low intensity. 
These ions could correspond to propylene (\ch{C3H6}), which is an expected product of DME conversion, and ketene (\ch{C2H2O}), which is a highly reactive intermediate of the reaction.
Due to moderate spectral resolution and low signal to noise ratio, it is not possible to distinguish them from the PES. 
However, it can be inferred 
that the signal corresponds to propylene, since ketene is so reactive that it is not expected to survive the transport through the second inert zone in the reactor. 
Furthermore, mass $14$ is the predominant fragment of ketene \cite{NIST}, which is not detected in our TOF spectrum (Fig. \ref{fig:map}c).
The comparison between extracted CIY-PES for mass $42$ and reference spectra for \ch{C3H6} and ketene is depicted in Fig. \ref{fig:CIY-PES}d. 

Ions with mass $44$ likely originate from \ch{CO2} and propane \ch{C3H8}, the latter being an expected minor product of DME conversion via hydrogen transfer reactions between propylene and methanol. 
However, CIY-PES of mass $44$ (Fig. S6) reveals that \ch{C3H8} does not exist in the chamber and all ions with mass $44$ are \ch{CO2+}.
We attribute the presence of \ch{CO2} to a combination of (i) the chamber background, (ii) the product of decarboxylation of MTH reaction intermediates on the zeolite \cite{huber2023}, as well as (iii) the product of oxidation of hydrocarbons by a minute amount of oxygen in the background (Fig. S5).

Ions with mass $56$ can be assigned to butene - another expected product of the reaction formed by methylation of propylene. 
However, the production of butene was limited, leading to low signal-to-noise ratio for CIY-PES of mass 56.

According to Cesarini et al. \cite{cesarini2022}, various isomers of \ch{C5H8} can undergo additional cyclization/dehydrogenation and methylation reactions, resulting in the formation of cyclopentadiene (\ch{C5H6}; m/z = $66$) and methyl cyclopentadiene (\ch{C6H8}; m/z = $80$), respectively. 
Furthermore, they have identified fulvene (\ch{C6H6}; m/z = $78$) formed through the dehydrogenation of methyl cyclopentadienes (\ch{C6H8}).
Fulvene is a precursor for the production of benzene, the first aromatic ring compound.  
Direct methylation of fulvene, as well as dehydrogenation-methylation of \ch{C6H8}, leads to the production of methyl fulvene (\ch{C7H8}; m/z = $92$), which is the primary precursor to toluene. 
This process can continue to generate other alkylated benzenes, including various isomers of xylene and trimethylbenzene.
In the current experiment, masses $78$ and $92$ are detected, while $66$ and $80$ are not observed (Fig. \ref{fig:map}b). 
To determine the identities of the detected peaks, a comparison between the CIY-PES obtained for masses $78$ and $92$, together with the relevant reference spectra are presented in Figs. \ref{fig:CIY-PES}e and \ref{fig:CIY-PES}f, respectively \cite{kimura1981, cesarini2022}. 
The results indicate that benzene and toluene are the parent molecules associated with the detected peaks. 
However, it was not possible to detect intermediates such as fulvene and methyl fulvene in our experiments. 
We hypothesize that this limitation could be attributed to the experimental configuration of the reactor used in this study, i.e., the second inert zone of the packed bed.

In the present study, two peaks at masses $105$ (\ch{C8H9+}) and $106$ (\ch{C8H10+}) are observed in the TOF spectrum as shown in Fig. \ref{fig:map}b. 
Considering the reaction mechanism, it is inferred that xylene (\ch{C8H10}) is the appropriate parent molecule associated with these peaks. 
The CIY-PES for mass $106$ and the aggregation of CIY-PES for xylenes main fragments (i.e., masses $106$, $105$, and $91$) are compared with reference spectra in Fig. \ref{fig:CIY-PES}g.
The CIY-PES for mass $106$ does not fully coincide with the reference spectra of xylenes \cite{koenig1974, salaneck1981} (see Fig. S7). 
However, the aggregation of CIY-PES for masses $106$, $105$, and $91$, presenting electrons coming from three various ionization channels, more closely matches the reference spectra.
This aligns with the practice in reference spectra, where electrons from all ionization channels (including both inner valence ionization and outermost shell ionization) are accounted for. 
Conversely, in CIY-PES for mass 106, only electrons produced in the ionization channel of
$\ch{C8H10}+h\nu \rightarrow \ch{C8H10+}+\ch{e-}$ are taken into consideration.
%

The reference spectra of xylenes shown in Fig. \ref{fig:CIY-PES}g are measured during a separate experiment without the reactor. 
Fig. \ref{fig:CIY-PES}g shows that the effluent stream contains a combination of different isomers of \ch{C8H10}, with m-xylene appearing to be the most abundant one.
Xylene isomerism offers a convenient benchmark problem for quantitative isomer discrimination, described in detail in section \ref{sec;Quan}, which is a novel aspect in PEPICO analysis introduced in our work.

In addition to the various isomers of xylene, isomers of trimethylbenzene can be generated through direct methylation or dehydrogenation-methylation of lighter hydrocarbons \cite{cesarini2022}.
Despite the detection of only a negligible number of cations with mass $120$ in the TOF spectrum (Fig. \ref{fig:map}b), the comparison of the extracted CIY-PES with reference spectra (Fig. \ref{fig:CIY-PES}h) indicates the presence of a mixture of different isomers of trimethylbenzenes in our experiment \cite{longetti2020}.
It is worthwhile to mention that, CIY-PES for mass 120 only shows the electrons coming from the ionization channel that leads to remove one electron. 
On the other hand, reference PES includes electrons coming from all ionization channels.


\subsection{Isomer quantification}\label{sec;Quan}
Quantitative discrimination of isomers from PEPICO data provides a valuable tool for analyzing complex reaction pathways in catalytic reactions of hydrocarbons. 
The reference spectra for different isomers of xylene, measured in the absence of reactor, are depicted in Fig. \ref{fig:CIY-PES}g along with the CIY-PES for mass 106 and its combination with masses $105$, and $91$ from the reactor effluent stream. 
The major differences between these spectra lies in their first and third bands in the $8-10$ and $13-14$ eV regions of binding energy, respectively.
The ionization channel of $\ch{C8H10}+h\nu \rightarrow \ch{C8H10+}+\ch{e-}$ is correlated to the electrons with binding energy of $8-10$ eV.
Moreover, for the reference spectra, electrons coming from the background (\ch{H2O}, \ch{N2}, and \ch{O2}) do not overlap with xylene electrons coming from ionization channel of \ch{C8H10+}.
Therefore, we are going to focus on the first band to quantitatively distinguish the ratio of different isomers of xylene in the effluent.
 
The comparison between CIY-PES for mass $106$ with reference spectra of xylene isomers is given in Fig. \ref{fig:Xylene_fit01}a.
The CIY-PES for mass $106$ does not have the two separate peaks which are characteristic for p-xylene. 
In addition, its FWHM is almost equal to m-xylene.
Therefore, m-xylene is the dominant isomer among the products. 
But what is the exact branching ratio?

In order to quantitatively determine the ratio of isomers, each reference spectrum has been deconvoluted using a collection of asymmetrically distorted Voigt profiles.
For reference PES of m- and p-xylene, aggregation of three Voigt profiles is required to adequately describe the collected data (Fig. \ref{fig:Xylene_fit01}b, c, respectively).
\begin{equation}
    PES_{m-xylene}=\omega^{m}_1 \cdot V^{m}_{1}+\omega^{m}_2 \cdot V^{m}_{2}+\omega^{m}_3 \cdot V^{m}_{3}
\end{equation}
\begin{equation}
    PES_{p-xylene}=\omega^{p}_1 \cdot V^{p}_{1}+\omega^{p}_2 \cdot V^{p}_{2}+\omega^{p}_3 \cdot V^{p}_{3}
\end{equation}
here, $\omega$ shows the individual weight of each Voigt function.
Then, these two groups of peaks were mixed in a specific ratio, while the peak shapes, center distances, and intensity ratios within each group were fixed according to their values estimated from pure reference compounds. 
By iteratively adjusting the ratio of these two groups of peaks and minimizing the difference between the generated spectra and the one with unknown mixing ratio, one could quantify the mixing ratio.
First, the method was benchmarked against two control mixtures of m- and p-xylenes with known compositions, 50:50 and 75:25, returning estimates of 56:44 and 77:23, respectively. 
This outcome demonstrates that, although promising, the isomer quantification analysis requires PEPICO data with higher signal to noise ratio, than in the present data, and more systematic collection of calibration datasets. 
Next, we applied the same routine to analyze the reactor effluent. 
Fig. \ref{fig:Xylene_fit01}d demonstrates the quality of the resulting model fit. 
Based on this method, and considering the area of peaks, almost 85\% of the electrons detected in coincidence with \ch{C8H10+} cation are coming from m-xylene and the remaining 15\% are related to p-xylene. 
This result agrees well with the selectivities reported in the literature for unmodified ZSM-5 \cite{zhang2015i, gao2020}.
Although p-xylene is more likely to be the prevalent primary product of toluene methylation inside the micropores, unmodified ZSM-5 in these studies is thought to contain external acid sites that rapidly isomerize p-xylene.
Unneberg and Kolboe \cite{unneberg1988} have shown that the xylene isomer distribution over ZSM-5 catalysts can shift from meta to para with increasing time on stream. 
In our future work, we will apply isomer-selective PEPICO to systematically characterize p-/m-xylene selectivity in zeolites at low-pressure conditions, i.e. in the limit of low DME exposure, to fully understand what controls their intrinsic xylene selectivity.

\section{Conclusion}
\textit{Operando} Photoelectron–Photoion Coincidence (PEPICO) mass-spectrometry is emerging as a valuable analytical tool for investigations of reaction mechanisms and kinetics in heterogeneous catalysis. Currently, the scope of the technique and its adoption in the catalysis research community are constrained by the limited availability of dedicated facilities around the globe. 
We have established the analysis of catalytic reactor effluents using PEPICO and, potentially, other photoionization-based methods at the FinEstBeAMS beamline at MAX-IV Laboratory.
This capability was demonstrated using dimethyl ether conversion on a prototypical ZSM-5 catalyst (at 375 \textdegree C and $ 5 \times 10^{-7}$ mbar total pressure) as a benchmark reaction, which agreed with the product distribution expected from the literature. 
Due to the specific configuration of the reactor packing in our proof-of-principle study, we have not observed either ketene or methyl radical, both highly-reactive intermediates that were recently revealed by \textit{operando} PEPICO. 
However, we have quantitatively determined the ratio of xylene isomers in the product stream by deconvoluting their coincidence photoelectron spectra, opening up a new avenue for quantitative isomer-selective PEPICO analysis in kinetic studies of heterogeneously catalyzed reactions. 

To enhance the technique, a differential pumping system is developing to facilitate high-pressure studies as well.
In the current study, the reactor pressure was in the range of $10^{-7}$-$10^{-6}$ mbar. 
However, with the implementation of the differential pumping system, it will be feasible to conduct PEPICO experiments while the reactor is in atmospheric pressure. 
Overcoming the constraint of being limited to steady-state conditions, due to prolonged data acquisition times, requires using different type of electron analyser, such as electron TOF, which has higher transmission than the hemispherical one.

\section{Acknowledgements}
This research was funded by the Research Council of Norway (\#272266, TAPXPS project), the Research Council of Finland (\#341288, IntriCAT project), and the University of Oulu Graduate School. We acknowledge MAX IV Laboratory for time on FinEstBeAMS Beamline under Proposals 20190444 and 20180449. Research conducted at MAX IV, a Swedish national user facility, is supported by the Swedish Research council under contract 2018-07152, the Swedish Governmental Agency for Innovation Systems under contract 2018-04969, and Formas under contract 2019-02496. We express our gratitude to Dr. Antti Kivimäki and Dr. Kirill Chernenko at the FinEstBeAMS beamline, MAX IV Laboratory for their assistance during the experiments.

\clearpage
\pagebreak

\bibliographystyle{unsrtnat}
\bibliography{iucr}

\begin{thebibliography}{38}
\providecommand{\natexlab}[1]{#1}
\providecommand{\url}[1]{\texttt{#1}}
\expandafter\ifx\csname urlstyle\endcsname\relax
  \providecommand{\doi}[1]{doi: #1}\else
  \providecommand{\doi}{doi: \begingroup \urlstyle{rm}\Url}\fi

\bibitem[Olsbye et~al.(2012)Olsbye, Svelle, Bj{\o}rgen, Beato, Janssens, Joensen, Bordiga, and Lillerud]{olsbye2012}
Unni Olsbye, Stian Svelle, Morten Bj{\o}rgen, Pablo Beato, Ton~VW Janssens, Finn Joensen, Silvia Bordiga, and Karl~Petter Lillerud.
\newblock Conversion of methanol to hydrocarbons: how zeolite cavity and pore size controls product selectivity.
\newblock \emph{Angewandte Chemie International Edition}, 51\penalty0 (24):\penalty0 5810--5831, 2012.

\bibitem[Hemberger et~al.(2020)Hemberger, van Bokhoven, P{\'e}rez-Ram{\'\i}rez, and Bodi]{hemberger2020new}
Patrick Hemberger, Jeroen~A van Bokhoven, Javier P{\'e}rez-Ram{\'\i}rez, and Andras Bodi.
\newblock New analytical tools for advanced mechanistic studies in catalysis: photoionization and photoelectron photoion coincidence spectroscopy.
\newblock \emph{Catalysis Science \& Technology}, 10\penalty0 (7):\penalty0 1975--1990, 2020.

\bibitem[Cejka et~al.(2010)Cejka, Corma, and Zones]{cejka2010}
Jiri Cejka, A.~Corma, and S.~Zones.
\newblock \emph{Zeolites and Catalysis: Synthesis, Reactions and Applications}.
\newblock 06 2010.
\newblock ISBN 9783527325146.
\newblock \doi{10.1002/9783527630295}.

\bibitem[Li and Yu(2021)]{li2021}
Yi~Li and Jihong Yu.
\newblock Emerging applications of zeolites in catalysis, separation and host-guest assembly.
\newblock \emph{Nature Reviews Materials}, 6\penalty0 (12):\penalty0 1156--1174, 2021.
\newblock \doi{10.1038/s41578-021-00347-3}.

\bibitem[Brogaard et~al.(2014)Brogaard, Henry, Schuurman, Medford, Moses, Beato, Svelle, N{\o}rskov, and Olsbye]{brogaard2014}
Rasmus~Y Brogaard, Reynald Henry, Yves Schuurman, Andrew~J Medford, Poul~Georg Moses, Pablo Beato, Stian Svelle, Jens~K N{\o}rskov, and Unni Olsbye.
\newblock Methanol-to-hydrocarbons conversion: The alkene methylation pathway.
\newblock \emph{Journal of catalysis}, 314:\penalty0 159--169, 2014.
\newblock \doi{10.1016/j.jcat.2014.04.006}.

\bibitem[Batchu et~al.(2017)Batchu, Galvita, Alexopoulos, Van~der Borght, Poelman, Reyniers, and Marin]{batchu2017}
Rakesh Batchu, Vladimir~V Galvita, Konstantinos Alexopoulos, Kristof Van~der Borght, Hilde Poelman, Marie-Fran{\c{c}}oise Reyniers, and Guy~B Marin.
\newblock Role of intermediates in reaction pathways from ethene to hydrocarbons over h-zsm-5.
\newblock \emph{Applied Catalysis A: General}, 538:\penalty0 207--220, 2017.
\newblock ISSN 0926-860X.
\newblock \doi{https://doi.org/10.1016/j.apcata.2017.03.013}.
\newblock URL \url{https://www.sciencedirect.com/science/article/pii/S0926860X17301126}.

\bibitem[Omojola et~al.(2021)Omojola, Logsdail, Van~Veen, and Nastase]{omojola2021}
Toyin Omojola, Andrew~J Logsdail, Andr{\'e}~C Van~Veen, and Stefan Adrian~F Nastase.
\newblock A quantitative multiscale perspective on primary olefin formation from methanol.
\newblock \emph{Physical Chemistry Chemical Physics}, 23\penalty0 (38):\penalty0 21437--21469, 2021.
\newblock \doi{10.1039/D1CP02551A}.
\newblock URL \url{http://dx.doi.org/10.1039/D1CP02551A}.

\bibitem[Redekop et~al.(2020)Redekop, Lazzarini, Bordiga, and Olsbye]{redekop2020temporal}
Evgeniy~A Redekop, Andrea Lazzarini, Silvia Bordiga, and Unni Olsbye.
\newblock A temporal analysis of products (tap) study of c2-c4 alkene reactions with a well-defined pool of methylating species on zsm-22 zeolite.
\newblock \emph{Journal of catalysis}, 385:\penalty0 300--312, 2020.

\bibitem[Kooser et~al.(2020)Kooser, Kivim{\"a}ki, Turunen, P{\"a}rna, Reisberg, Kirm, Valden, Huttula, and Kukk]{kooser2020gas}
Kuno Kooser, Antti Kivim{\"a}ki, Paavo Turunen, Rainer P{\"a}rna, Liis Reisberg, Marco Kirm, Mika Valden, Marko Huttula, and Edwin Kukk.
\newblock Gas-phase endstation of electron, ion and coincidence spectroscopies for diluted samples at the finestbeams beamline of the max iv 1.5 gev storage ring.
\newblock \emph{Journal of Synchrotron Radiation}, 27\penalty0 (4):\penalty0 1080--1091, 2020.

\bibitem[Arion and Hergenhahn(2015)]{arion2015coincidence}
Tiberiu Arion and Uwe Hergenhahn.
\newblock Coincidence spectroscopy: Past, present and perspectives.
\newblock \emph{Journal of Electron Spectroscopy and Related Phenomena}, 200:\penalty0 222--231, 2015.

\bibitem[Wen et~al.(2020)Wen, Yu, Zhou, Ma, Zhou, Cao, Yang, Xu, Qi, Zhang, et~al.]{wen2020}
Wu~Wen, Shengsheng Yu, Chaoqun Zhou, Hao Ma, Zhongyue Zhou, Chuangchuang Cao, Jiuzhong Yang, Minggao Xu, Fei Qi, Guobin Zhang, et~al.
\newblock Formation and fate of formaldehyde in methanol-to-hydrocarbon reaction: In situ synchrotron radiation photoionization mass spectrometry study.
\newblock \emph{Angewandte Chemie International Edition}, 59\penalty0 (12):\penalty0 4873--4878, 2020.

\bibitem[Cesarini et~al.(2022)Cesarini, Mitchell, Zichittella, Agrachev, Schmid, Jeschke, Pan, Bodi, Hemberger, and P{\'e}rez-Ram{\'\i}rez]{cesarini2022}
Alessia Cesarini, Sharon Mitchell, Guido Zichittella, Mikhail Agrachev, Stefan~P Schmid, Gunnar Jeschke, Zeyou Pan, Andras Bodi, Patrick Hemberger, and Javier P{\'e}rez-Ram{\'\i}rez.
\newblock Elucidation of radical-and oxygenate-driven paths in zeolite-catalysed conversion of methanol and methyl chloride to hydrocarbons.
\newblock \emph{Nature Catalysis}, pages 1--10, 2022.

\bibitem[SLS(2023)]{SLS}
SLS.
\newblock The vuv beamline, 2023.
\newblock URL \url{https://www.psi.ch/en/sls/vuv}.

\bibitem[Zhou et~al.(2016)Zhou, Du, Yang, Wang, Li, Wei, Du, Li, Qi, and Wang]{NSRL2016}
Zhongyue Zhou, Xuewei Du, Jiuzhong Yang, Yizun Wang, Chaoyang Li, Shen Wei, Liangliang Du, Yuyang Li, Fei Qi, and Qiuping Wang.
\newblock The vacuum ultraviolet beamline/endstations at nsrl dedicated to combustion research.
\newblock \emph{Journal of synchrotron radiation}, 23\penalty0 (4):\penalty0 1035--1045, 2016.

\bibitem[Tang et~al.(2015)Tang, Garcia, Gil, and Nahon]{tang2015}
Xiaofeng Tang, Gustavo~A Garcia, Jean-Fran{\c{c}}ois Gil, and Laurent Nahon.
\newblock Vacuum upgrade and enhanced performances of the double imaging electron/ion coincidence end-station at the vacuum ultraviolet beamline desirs.
\newblock \emph{Review of Scientific Instruments}, 86\penalty0 (12), 2015.

\bibitem[Xie and Olsbye(2023)]{xie2023}
Jingxiu Xie and Unni Olsbye.
\newblock The oxygenate-mediated conversion of cox to hydrocarbons- on the role of zeolites in tandem catalysis.
\newblock \emph{Chemical reviews}, 123\penalty0 (20):\penalty0 11775--11816, 2023.
\newblock \doi{10.1021/acs.chemrev.3c00058}.

\bibitem[P{\"a}rna et~al.(2017)P{\"a}rna, Sankari, Kukk, N{\~o}mmiste, Valden, Lastusaari, Kooser, Kokko, Hirsim{\"a}ki, Urpelainen, et~al.]{parna2017}
Rainer P{\"a}rna, Rami Sankari, Edwin Kukk, E~N{\~o}mmiste, Mika Valden, Marko Lastusaari, Kuno Kooser, Kalevi Kokko, Mika Hirsim{\"a}ki, Samuli Urpelainen, et~al.
\newblock Finestbeams--a wide-range finnish-estonian beamline for materials science at the 1.5 gev storage ring at the max iv laboratory.
\newblock \emph{Nuclear Instruments and Methods in Physics Research Section A: Accelerators, Spectrometers, Detectors and Associated Equipment}, 859:\penalty0 83--89, 2017.

\bibitem[Kukk et~al.(2007)Kukk, Sankari, Huttula, Sankari, Aksela, and Aksela]{kukk2007}
E~Kukk, R~Sankari, M~Huttula, A~Sankari, H~Aksela, and S~Aksela.
\newblock New electron-ion coincidence setup: Fragmentation of acetonitrile following n 1s core excitation.
\newblock \emph{Journal of electron spectroscopy and related phenomena}, 155\penalty0 (1-3):\penalty0 141--147, 2007.

\bibitem[Chernenko et~al.(2021)Chernenko, Kivim{\"a}ki, P{\"a}rna, Wang, Sankari, Leandersson, Tarawneh, Pankratov, Kook, Kukk, et~al.]{chernenko2021}
Kirill Chernenko, Antti Kivim{\"a}ki, Rainer P{\"a}rna, Weimin Wang, Rami Sankari, Mats Leandersson, Hamed Tarawneh, Vladimir Pankratov, Mati Kook, Edwin Kukk, et~al.
\newblock Performance and characterization of the finestbeams beamline at the max iv laboratory.
\newblock \emph{Journal of Synchrotron Radiation}, 28\penalty0 (5), 2021.

\bibitem[Rojo-Gama et~al.(2018)Rojo-Gama, Mentel, Kalantzopoulos, Pappas, Dovgaliuk, Olsbye, Lillerud, Beato, Lundegaard, Wragg, et~al.]{rojo2018}
Daniel Rojo-Gama, Lukasz Mentel, Georgios~N Kalantzopoulos, Dimitrios~K Pappas, Iurii Dovgaliuk, Unni Olsbye, Karl~Petter Lillerud, Pablo Beato, Lars~F Lundegaard, David~S Wragg, et~al.
\newblock Deactivation of zeolite catalyst h-zsm-5 during conversion of methanol to gasoline: operando time-and space-resolved x-ray diffraction.
\newblock \emph{The Journal of Physical Chemistry Letters}, 9\penalty0 (6):\penalty0 1324--1328, 2018.

\bibitem[Bodi et~al.(2013)Bodi, Hemberger, Osborn, and Szt\'aray]{bodi2013}
Andras Bodi, Patrick Hemberger, David~L Osborn, and B\'alint Szt\'aray.
\newblock Mass-resolved isomer-selective chemical analysis with imaging photoelectron photoion coincidence spectroscopy.
\newblock \emph{The Journal of Physical Chemistry Letters}, 4\penalty0 (17):\penalty0 2948--2952, 2013.

\bibitem[Pr{\"u}mper and Ueda(2007)]{prumper2007}
G~Pr{\"u}mper and K~Ueda.
\newblock Electron--ion--ion coincidence experiments for photofragmentation of polyatomic molecules using pulsed electric fields: Treatment of random coincidences.
\newblock \emph{Nuclear Instruments and Methods in Physics Research Section A: Accelerators, Spectrometers, Detectors and Associated Equipment}, 574\penalty0 (2):\penalty0 350--362, 2007.

\bibitem[Koenig et~al.(1974)Koenig, Tuttle, and Wielesek]{koenig1974}
T~Koenig, M~Tuttle, and RA~Wielesek.
\newblock The he (i) photoelectron spectra of xylenes and metacyclophanes. a reassignment of the lowest ionic state of [2.2] metacyclophane.
\newblock \emph{Tetrahedron Letters}, 15\penalty0 (29):\penalty0 2537--2540, 1974.

\bibitem[Ying et~al.(1993)Ying, Zhu, Mathers, Gover, Banjav{\v{c}}i{\'c}, Zheng, Brion, and Leung]{ying1993}
JF~Ying, H~Zhu, CP~Mathers, BN~Gover, MP~Banjav{\v{c}}i{\'c}, Y~Zheng, CE~Brion, and KT~Leung.
\newblock Electron-momentum-specific valence-shell electronic structures of cis-, trans-, and iso-butene by symmetric noncoplanar (e, 2 e) spectroscopy.
\newblock \emph{The Journal of chemical physics}, 98\penalty0 (6):\penalty0 4512--4519, 1993.

\bibitem[Wallace(2018)]{NIST}
William~E Wallace.
\newblock Mass spectra.
\newblock \emph{NIST chemistry webbook, NIST standard reference database}, \penalty0 (69):\penalty0 20899, 2018.

\bibitem[Feng et~al.(2001)Feng, Cooper, and Brion]{feng2001}
Renfei Feng, Glyn Cooper, and CE~Brion.
\newblock Ionic photofragmentation and photoionization of dimethyl ether in the vuv and soft x-ray regions (8.5--80 ev)--absolute oscillator strengths for molecular and dissociative photoionization.
\newblock \emph{Chemical Physics}, 270\penalty0 (2):\penalty0 319--332, 2001.

\bibitem[Chang(1983)]{Chang1983}
Clarence~D. Chang.
\newblock Hydrocarbons from methanol.
\newblock \emph{Catalysis Reviews}, 25\penalty0 (1):\penalty0 1--118, 1983.
\newblock \doi{10.1080/01614948308078874}.
\newblock URL \url{https://doi.org/10.1080/01614948308078874}.

\bibitem[Rohatgi(2022)]{Rohatgi2022}
Ankit Rohatgi.
\newblock Webplotdigitizer: Version 4.6, 2022.
\newblock URL \url{https://automeris.io/WebPlotDigitizer}.

\bibitem[Kimura(1981)]{kimura1981}
Katsumi Kimura.
\newblock \emph{Handbook of HeI photoelectron spectra of fundamental organic molecules}.
\newblock Halsted Press, 1981.

\bibitem[Chang et~al.(2017)Chang, Xiong, Bross, Ruscic, and Ng]{chang2017}
Yih-Chung Chang, Bo~Xiong, David~H Bross, Branko Ruscic, and CY~Ng.
\newblock A vacuum ultraviolet laser pulsed field ionization-photoion study of methane (ch 4): determination of the appearance energy of methylium from methane with unprecedented precision and the resulting impact on the bond dissociation energies of ch 4 and ch 4+.
\newblock \emph{Physical Chemistry Chemical Physics}, 19\penalty0 (14):\penalty0 9592--9605, 2017.

\bibitem[Stockbauer and Inghram(1975)]{stockbauer1975}
Roger Stockbauer and Mark~G Inghram.
\newblock Threshold photoelectron--photoion coincidence mass spectrometric study of ethylene and ethylene-d 4.
\newblock \emph{The Journal of Chemical Physics}, 62\penalty0 (12):\penalty0 4862--4870, 1975.

\bibitem[Huber and Plessow(2023)]{huber2023}
Philipp Huber and Philipp~N Plessow.
\newblock The role of decarboxylation reactions during the initiation of the methanol-to-olefins process.
\newblock \emph{Journal of Catalysis}, 428:\penalty0 115134, 2023.

\bibitem[Salaneck(1981)]{salaneck1981}
W.~R. Salaneck.
\newblock \emph{Intermolecular Relaxation Effects in the Ultraviolet Photoelectron Spectroscopy of Molecular Solids}, chapter~11, pages 121--149.
\newblock ACS Publications, 1981.
\newblock \doi{10.1021/bk-1981-0162.ch011}.
\newblock URL \url{https://pubs.acs.org/doi/abs/10.1021/bk-1981-0162.ch011}.

\bibitem[Longetti et~al.(2020)Longetti, Randulov{\'a}, Ojeda, Mewes, Miseikis, Grilj, Sanchez-Gonzalez, Witting, Siegel, Diveki, et~al.]{longetti2020}
L~Longetti, M~Randulov{\'a}, J~Ojeda, L~Mewes, L~Miseikis, J~Grilj, A~Sanchez-Gonzalez, T~Witting, T~Siegel, Z~Diveki, et~al.
\newblock Photoemission from non-polar aromatic molecules in the gas and liquid phase.
\newblock \emph{Physical Chemistry Chemical Physics}, 22\penalty0 (7):\penalty0 3965--3974, 2020.

\bibitem[Zhang et~al.(2015)Zhang, Qian, Kong, and Wei]{zhang2015i}
Jingui Zhang, Weizhong Qian, Chuiyan Kong, and Fei Wei.
\newblock Increasing para-xylene selectivity in making aromatics from methanol with a surface-modified zn/p/zsm-5 catalyst.
\newblock \emph{ACS Catalysis}, 5\penalty0 (5):\penalty0 2982--2988, 2015.

\bibitem[Gao et~al.(2020)Gao, Guo, Cui, Yang, He, Zeng, Taguchi, Abe, Ma, Yoneyama, et~al.]{gao2020}
Weizhe Gao, Lisheng Guo, Yu~Cui, Guohui Yang, Yingluo He, Chunyang Zeng, Akira Taguchi, Takayuki Abe, Qingxiang Ma, Yoshiharu Yoneyama, et~al.
\newblock Selective conversion of co2 into para-xylene over a zncr2o4-zsm-5 catalyst.
\newblock \emph{ChemSusChem}, 13\penalty0 (24):\penalty0 6541--6545, 2020.

\bibitem[Unneberg and Kolboe(1988)]{unneberg1988}
Erik Unneberg and Stein Kolboe.
\newblock Formation of p-xylene from methanol over h-zsm-5.
\newblock In \emph{Studies in Surface Science and Catalysis}, volume~36, pages 195--199. Elsevier, 1988.

\bibitem[Szt{\'a}ray et~al.(2017)Szt{\'a}ray, Voronova, Torma, Covert, Bodi, Hemberger, Gerber, and Osborn]{sztaray2017crf}
B{\'a}lint Szt{\'a}ray, Krisztina Voronova, Kriszti{\'a}n~G Torma, Kyle~J Covert, Andras Bodi, Patrick Hemberger, Thomas Gerber, and David~L Osborn.
\newblock Crf-pepico: Double velocity map imaging photoelectron photoion coincidence spectroscopy for reaction kinetics studies.
\newblock \emph{The Journal of chemical physics}, 147\penalty0 (1), 2017.

\end{thebibliography}

\pagebreak


\begin{figure}
\caption{The schematic figure of the reactor connected to the gas-phase endstation at the FinEstBeAMS beamline of the MAX IV Laboratory.}
\includegraphics[width=1\textwidth]{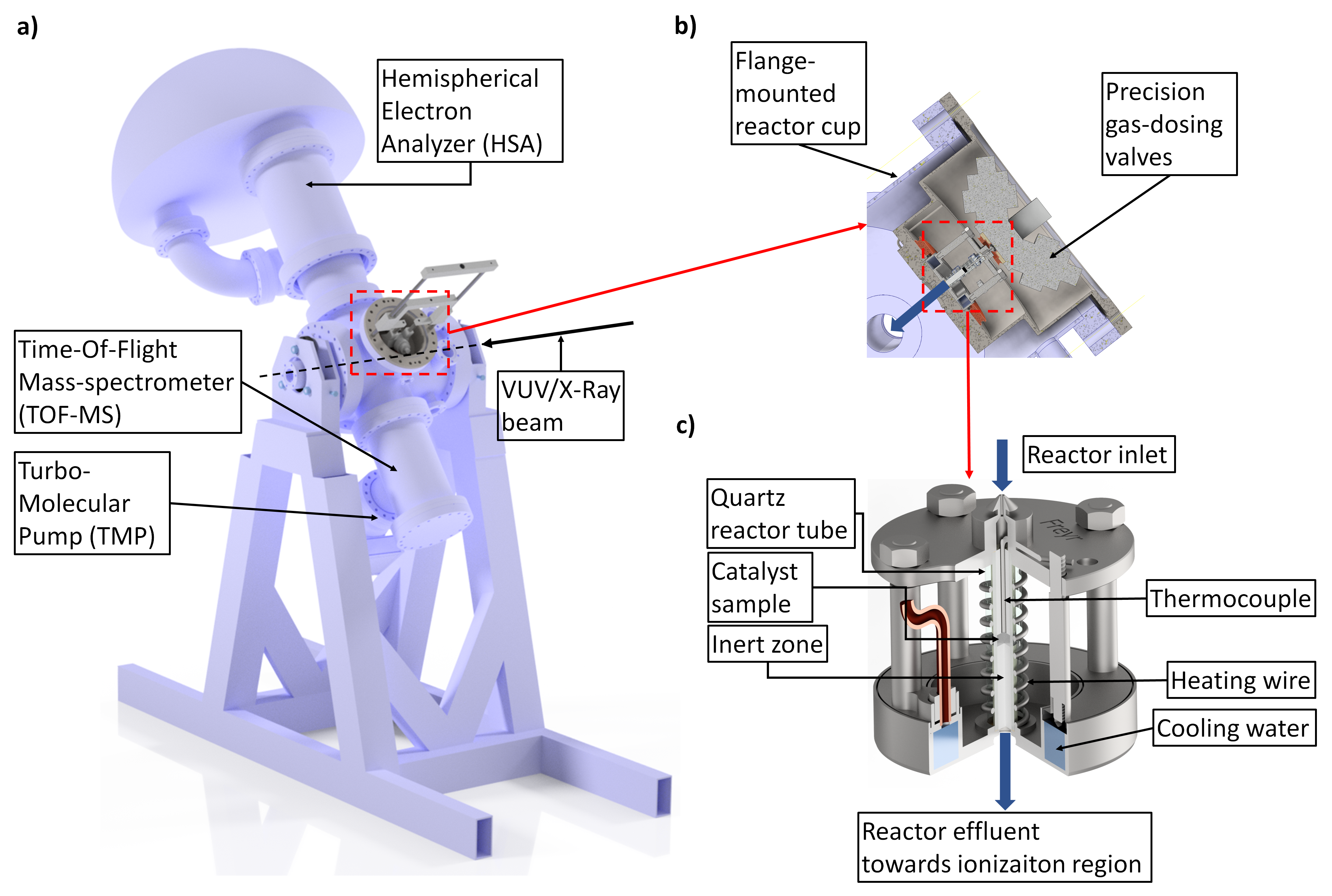}
\label{fig:GPES}
\end{figure}

\begin{figure}
\caption{a) PEPICO map of the effluent stream of DME conversion over ZSM5 
   zeolite at 375 \textdegree C ionized using light with 40 eV energy, 
   b) integrated binding energy spectrum. 
   , and c) integrated ion TOF spectrum with false coincidences subtracted. For higher TOF, the spectrum intensity is multiplied by 15, to make it more clear.
   The corresponding mass per charge ratios are mentioned for groups of TOF peaks.}
\includegraphics[width=\textwidth]{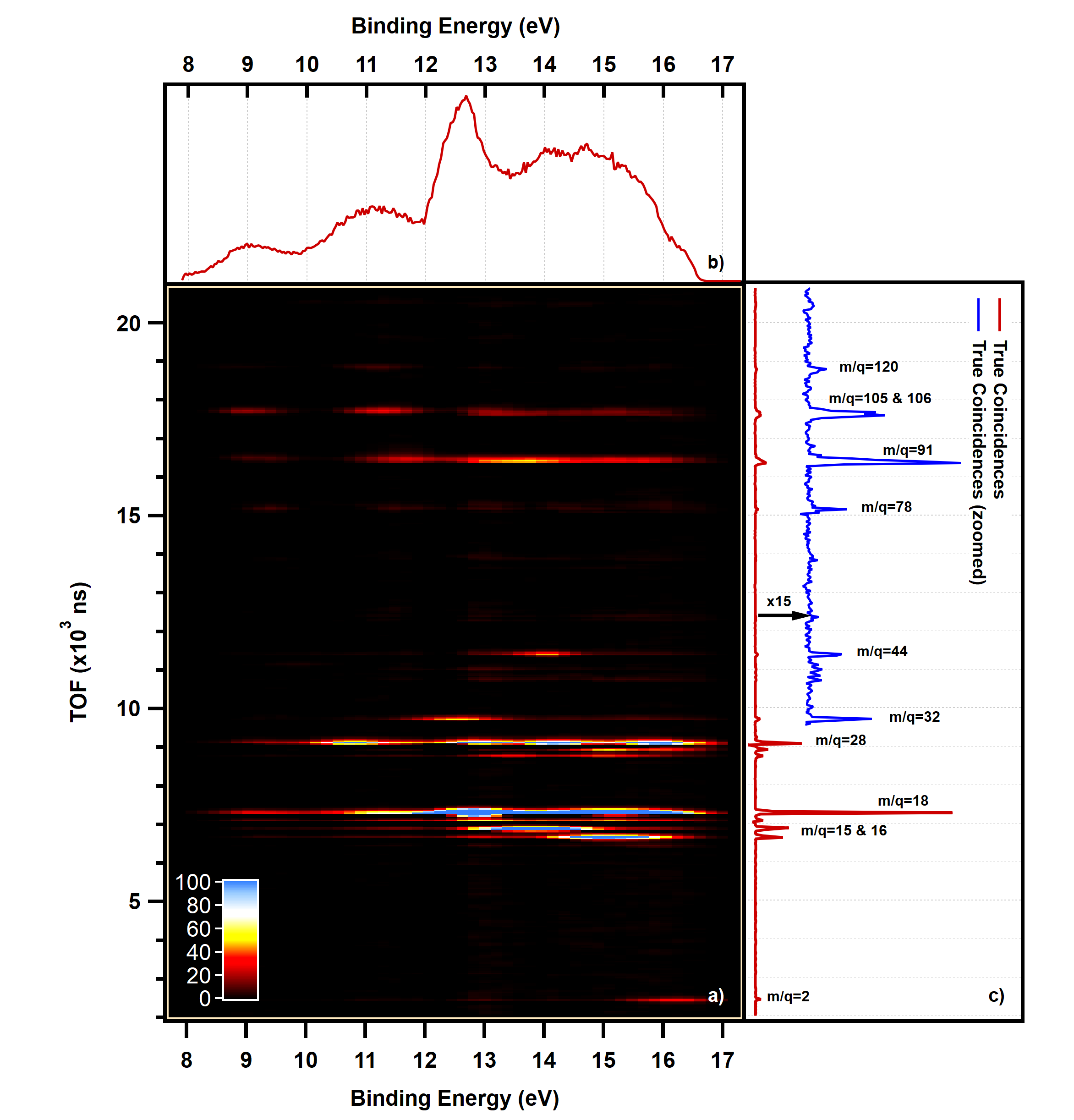}
\label{fig:map}
\end{figure}

\begin{figure}
\caption{The comparison of extracted coincidence ion yield photoelectron spectra (CIY-PES) at $40$ eV photon energy with reference spectra for the most intense detected cations \cite{kimura1981, cesarini2022, koenig1974, longetti2020}.}
\includegraphics[width=1\textwidth]{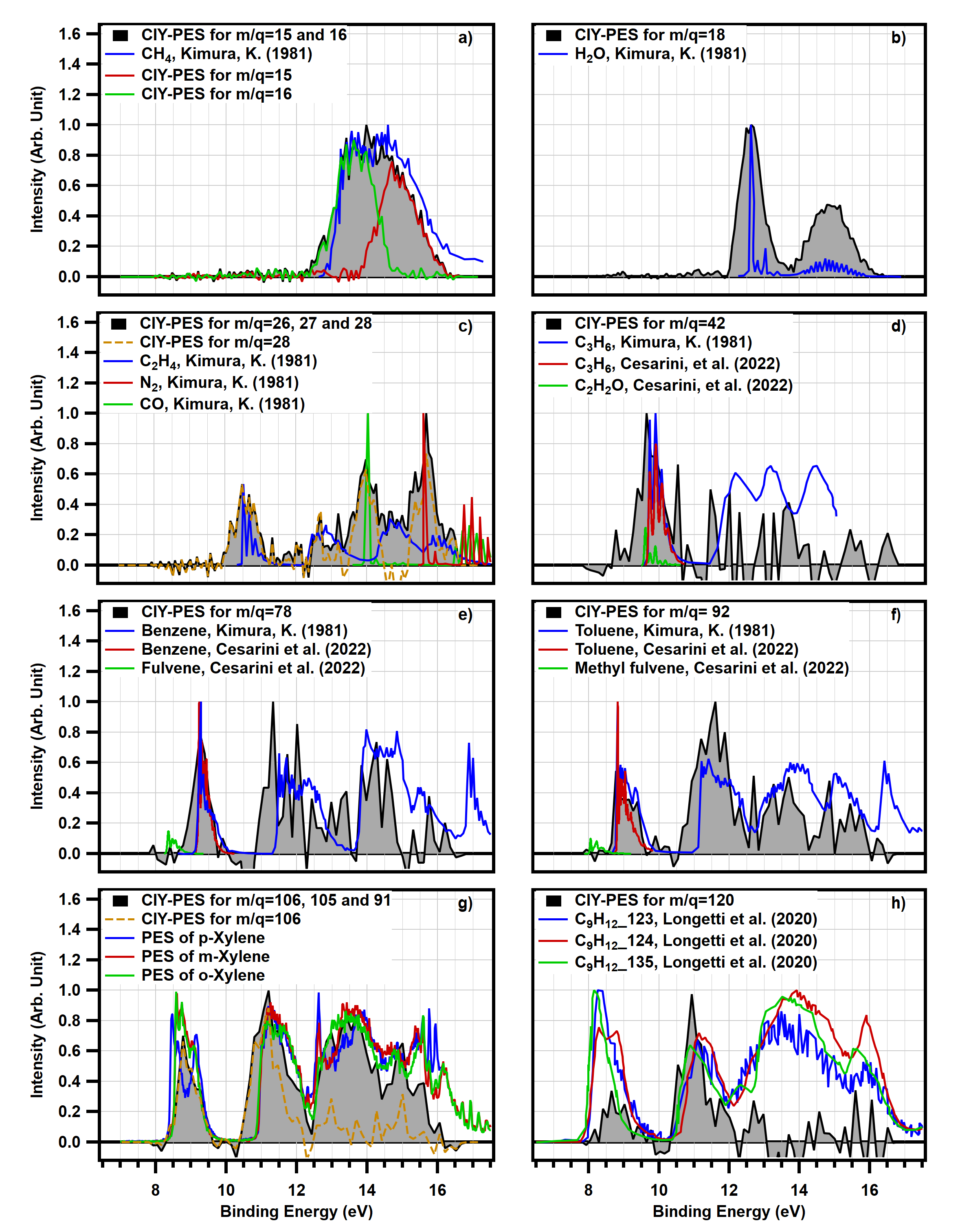}
\label{fig:CIY-PES}
\end{figure}

\begin{figure}
\caption{Comparison between a) CIY-PES for mass 106 with reference PES of m-xylene and p-xylene, b) and c) reference PES of m-xylene and p-xylene respectively, with their corresponding fits, and d) CIY-PES for mass 106 with the deconvoluted peaks.}
\includegraphics[width=1\textwidth]{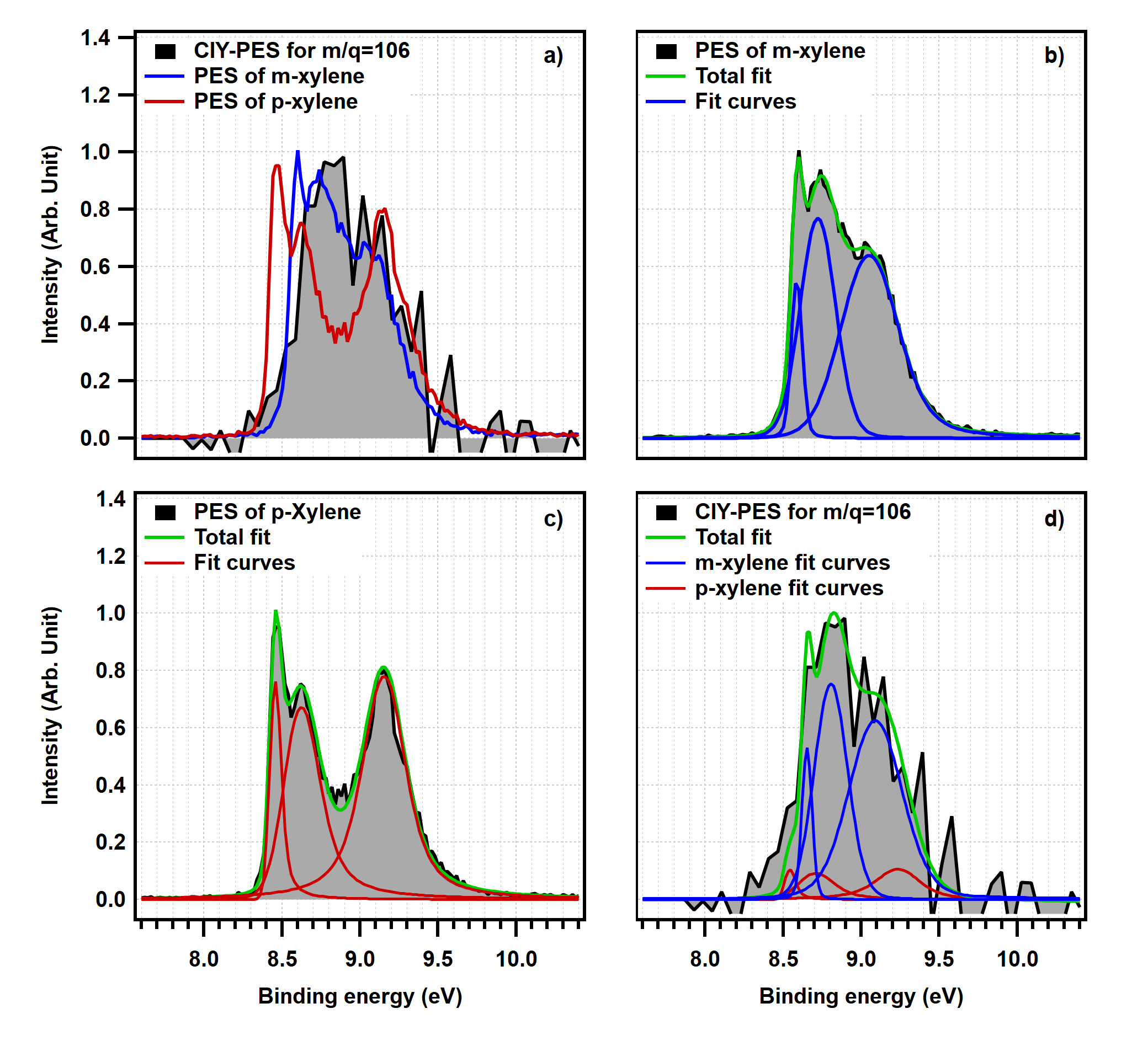}
\label{fig:Xylene_fit01}
\end{figure}

\pagebreak

\begin{table}[]
\caption{Catalytic experiments at FinEstBeAMS in the global context}
\label{tab:facilities}
\begin{tabular}{p{1.35cm}p{2cm}p{1.4cm}p{1.25cm}p{1.9cm}p{1.9cm}p{1.6cm}p{1.2cm}}
Facility &
  Beamline &
  Pressure range  &
  Energy range &
  Analyzer &
  Detector &
  Techniques &
  Ref. \\ \hline
SLS\textsuperscript{a} &
  VUV &
  $10^{-8}$–$10^3$ mbar &
  $3$–$150$ eV &
  e-TOF &
  position-sensitive DLD &
  PIMS iPEPICO i\textsuperscript{2}PEPICO &
  \cite{sztaray2017crf}, \\
  & & & & ion-TOF & position-sensitive DLD &  &
  and \cite{SLS} \\
  & & & & & & \\
NSRL &
  BL03U &
  $30$–$760$ Torr &
  $5$–$21$ eV &
  ion-TOF &
  cone-shaped stainless steel anode &
  PIMS &
  \cite{NSRL2016} \\
  & & & & & & &
  \\
SOLEIL &
  DESIRS (SAPHIRS endstation) &
  $10^{-8}$ mbar to a few mTorr &
  $5$–$40$ eV &
  e-TOF &
  position-sensitive detector &
  i\textsuperscript{2}PEPICO &
  \cite{tang2015} \\ 
   & & & & ion-TOF & position-sensitive detector & \\
   & & & & & & &
   \\
MAX IV &
  FinEstBeAMS (GPES) &
  $10^{-8}$–$10^{-6}$ mbar\textsuperscript{b} &
  $4.4$–$1000$ eV &
  hemispherical electron energy analyzer &
  position-sensitive resistive anode &
  XPS PIMS PEPICO PIPICO & 
  \cite{parna2017}, and \\ 
   & & & & ion-TOF & position-sensitive DLD & &
   \cite{kooser2020gas} \\
\multicolumn{8}{l}{}\\
\multicolumn{8}{l}{\textsuperscript{a} SLS shutdown in $2023$ will limit global PEPICO capacity for a few years.}\\
\multicolumn{8}{p{14.9cm}}{\textsuperscript{b} Differentially-pumped Molecular Beam Extraction (MBE) for experiments up to $10^3$ mbar is under design/construction at the University of Oulu.}
\end{tabular}
\end{table}

\begin{table}
\caption{Detailed breakdown of ionisation fragments at 40 eV photon energy (in coincidence with 7 to 17 eV binding energy), corresponding to the TOF spectrum of true coincidences}

\begin{tabular}{ccccc}  

 Mass (a.m.u.)  & Chemical formula  & \hspace{1cm} & Mass (a.m.u.)   & Chemical formula   \\
\hline
 1	   & \ch{H+}                       & & 39-42   & \ch{C3H_n+}   ($n=3-6$) \\
 2	   & \ch{H2+}                      & & 44	   & \ch{CO2+}               \\
14	    & \ch{N+}                       & & 50-52   & \ch{C4H_n+}  ($n=2-4$)  \\
15-16	& \ch{CH_n+}   ($n=3-4$)        & & 65	    & \ch{C5H5+}              \\
17    	& \ch{OH+}                      & & 77-78   & \ch{C6H_n+}  ($n=5-6$)  \\
18	    & \ch{H2O+}                     & & 91-92   & \ch{C7H_n+}  ($n=7-8$)  \\
26-27	& \ch{C2H_n+}   ($n=2-3$)       & & 105-106 & \ch{C8H_n+}  ($n=9-10$) \\
28	    & \ch{C2H4+}, \ch{N2+}, \ch{CO+}& & 120		& \ch{C9H12+}             \\
29-30	& \ch{C2H_n+}   ($n=5-6$)       & & 128		& \ch{C10H8+}             \\
32	    & \ch{O2+}                      & & 141-142 & \ch{C11H_n+} ($n=9-10$) \\
37	    & \ch{C3H+}                     & & 156		& \ch{C12H12+}            \\  
\label{table:ions}
\end{tabular}
\end{table}

\clearpage
\pagebreak

\appendix
\section{Supplementary information}

\textbf{\Large{PEPICO analysis of catalytic reactor effluents towards quantitative isomer discrimination: DME conversion over a ZSM-5 zeolite}}

\large{Morsal~Babayan,$^{a*}$ Evgeniy~Redekop,$^{b}$ Esko~Kokkonen,$^{c}$ Unni~Olsbye,$^{b}$ Marko~Huttula,$^{a}$ Samuli~Urpelainen,$^{a*}$}

$^{a}$ Nano and Molecular Systems Research Unit, University of Oulu, Oulu, Finland \\
$^{b}$ Department of Chemistry, Centre for Materials Science and Nanotechnology (SMN), University of Oslo, Oslo, Norway \\
$^{c}$ MAX IV Laboratory, Lund University, Lund, Sweden

$^{*}$ Corresponding author. E-mails: \texttt{morsal.babayan@oulu.fi, samuli.urpelainen@oulu.fi} \\


\vspace{16.5cm}

\begin{figure}
\centering
\includegraphics[width=1.1\textwidth]{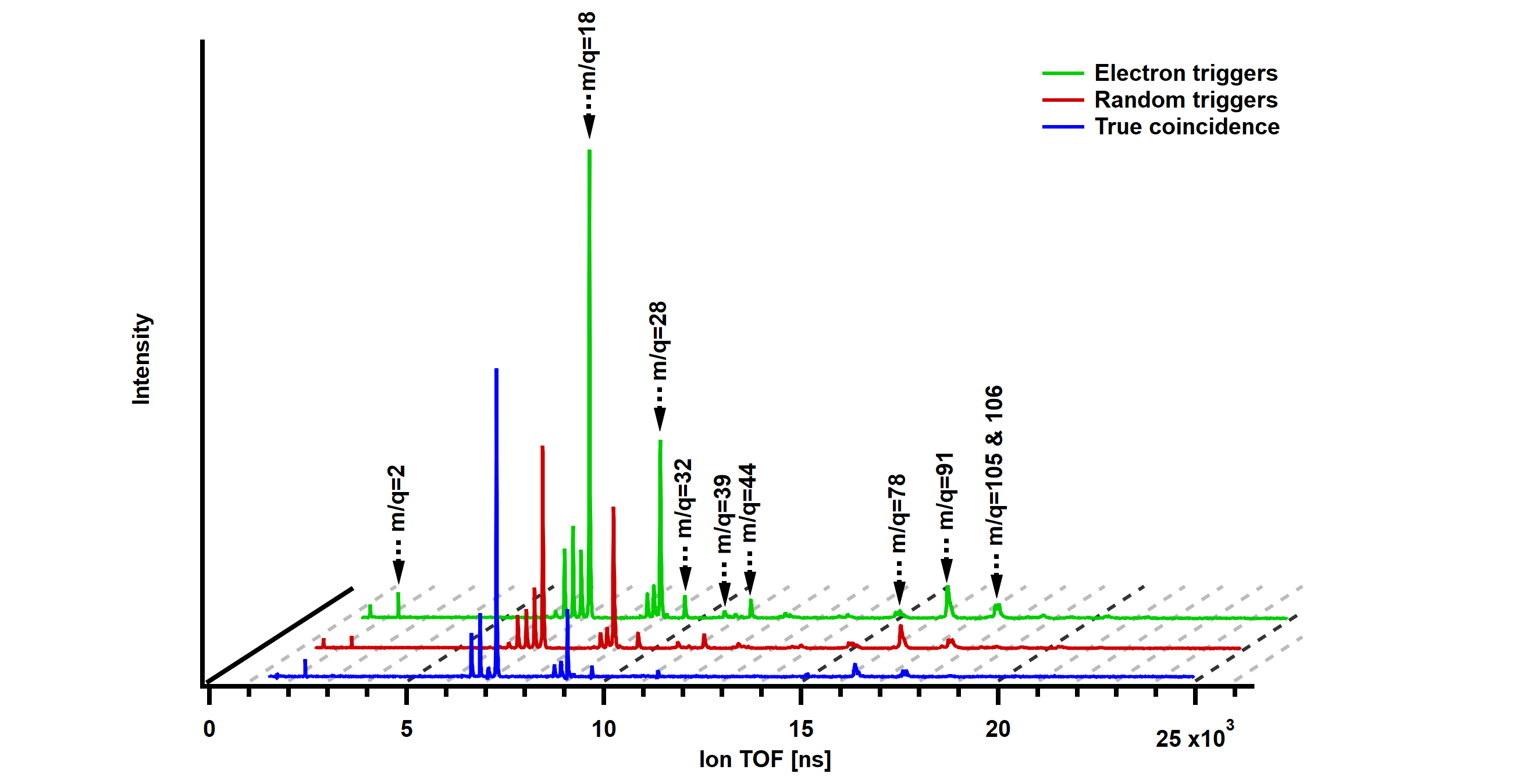}
\caption{The TOF spectra of ions for random triggers, electron triggers, and true coincidences recorded with photon energy of $40$ eV.}
\label{figS:TOF}
\end{figure}

\begin{figure}
\centering
\includegraphics[width=0.8\textwidth]{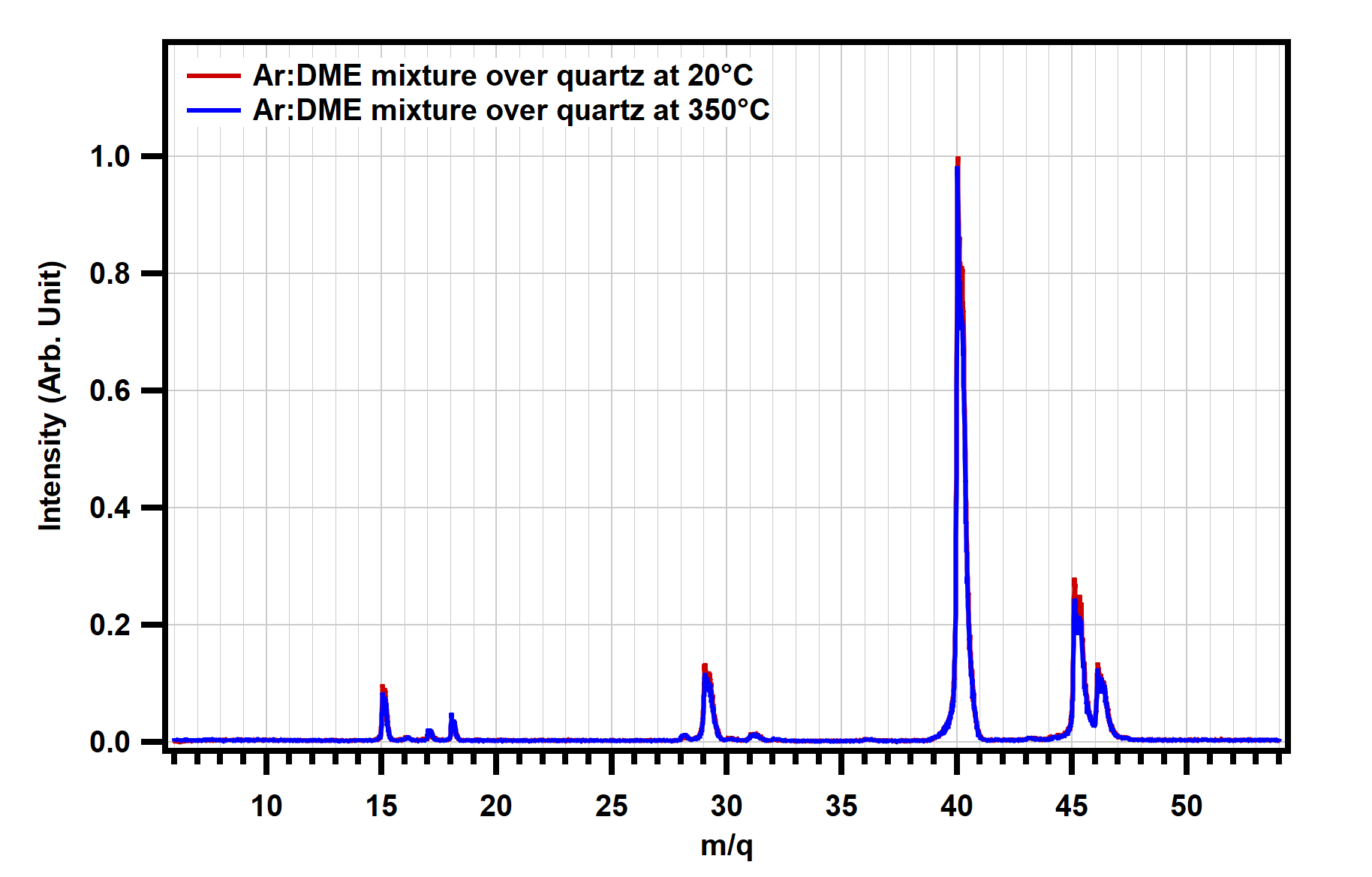}
\caption{The TOF spectra of a mixture of Ar and DME (3:1 ratio) flowing over the empty reactor at 20 and 350 \textdegree C temperature.}
\label{figS:blank}
\end{figure}

\begin{figure}
\centering
\includegraphics[width=0.8\textwidth]{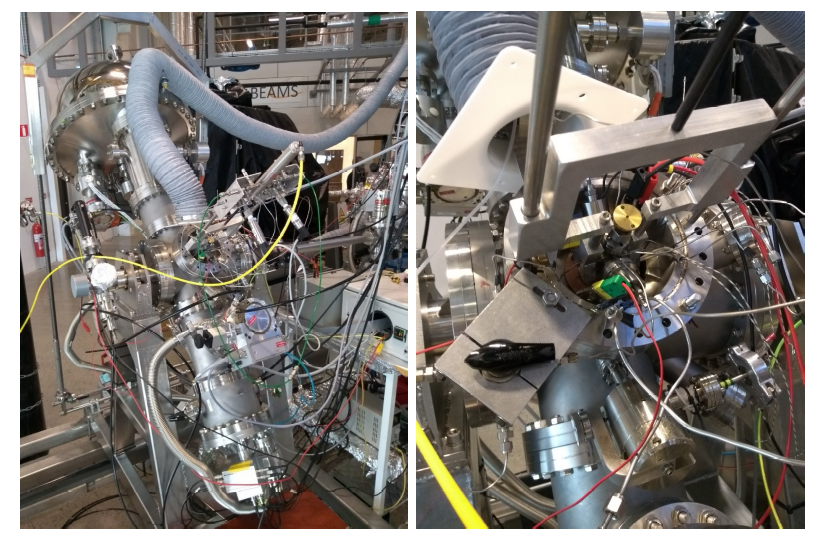}
\caption{Photographs of the experimental setup at the GPES.}
\label{figS:reactor}
\end{figure}

\begin{figure}
\centering
\includegraphics[width=0.70\textwidth]{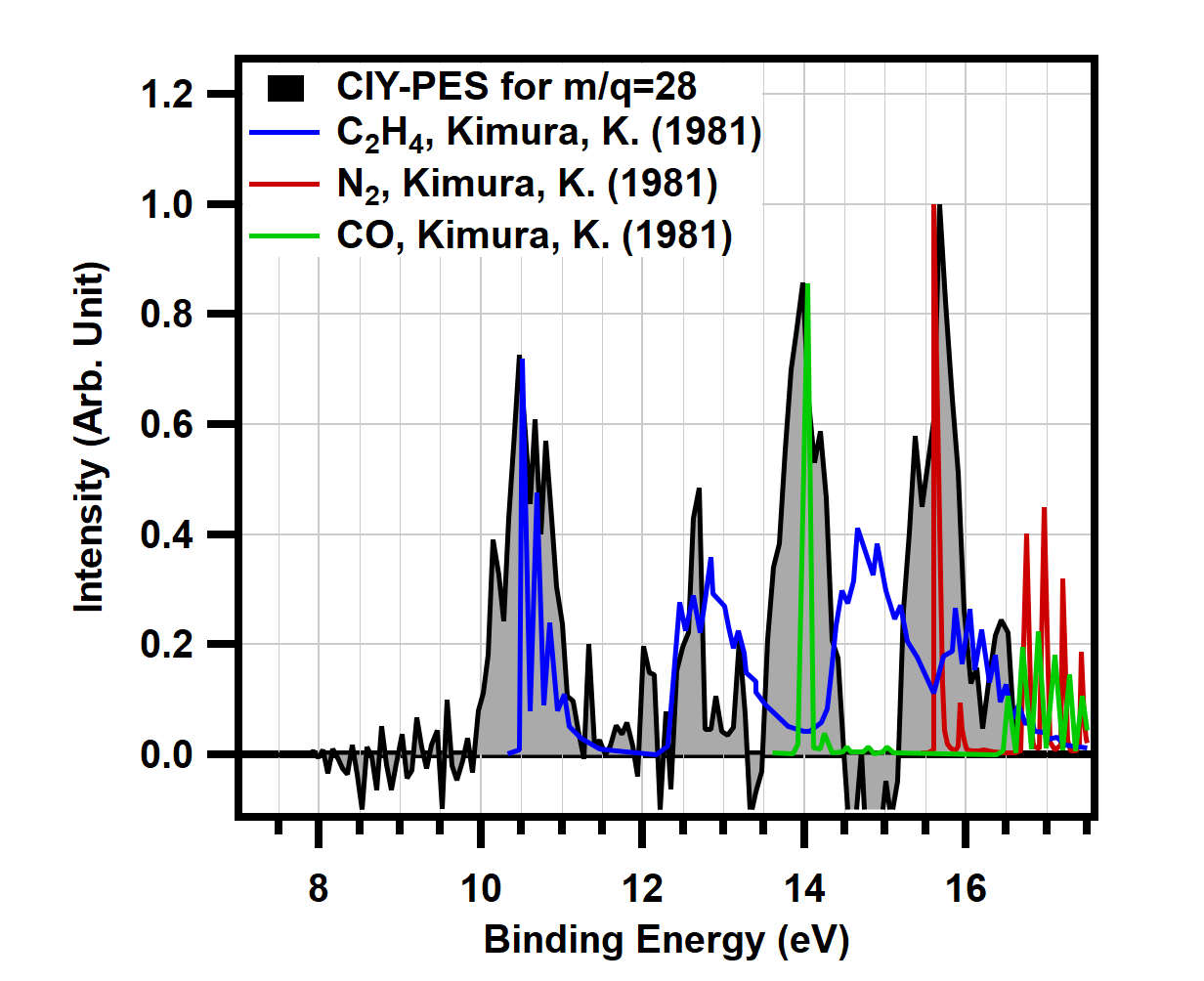}
\caption{The comparison of extracted coincidence ion yield PES at $40$ eV photon energy with reference spectra for ions with mass $28$ \cite{kimura1981}.}
\label{figS:28}
\end{figure}

\begin{figure}
\centering
\includegraphics[width=0.70\textwidth]{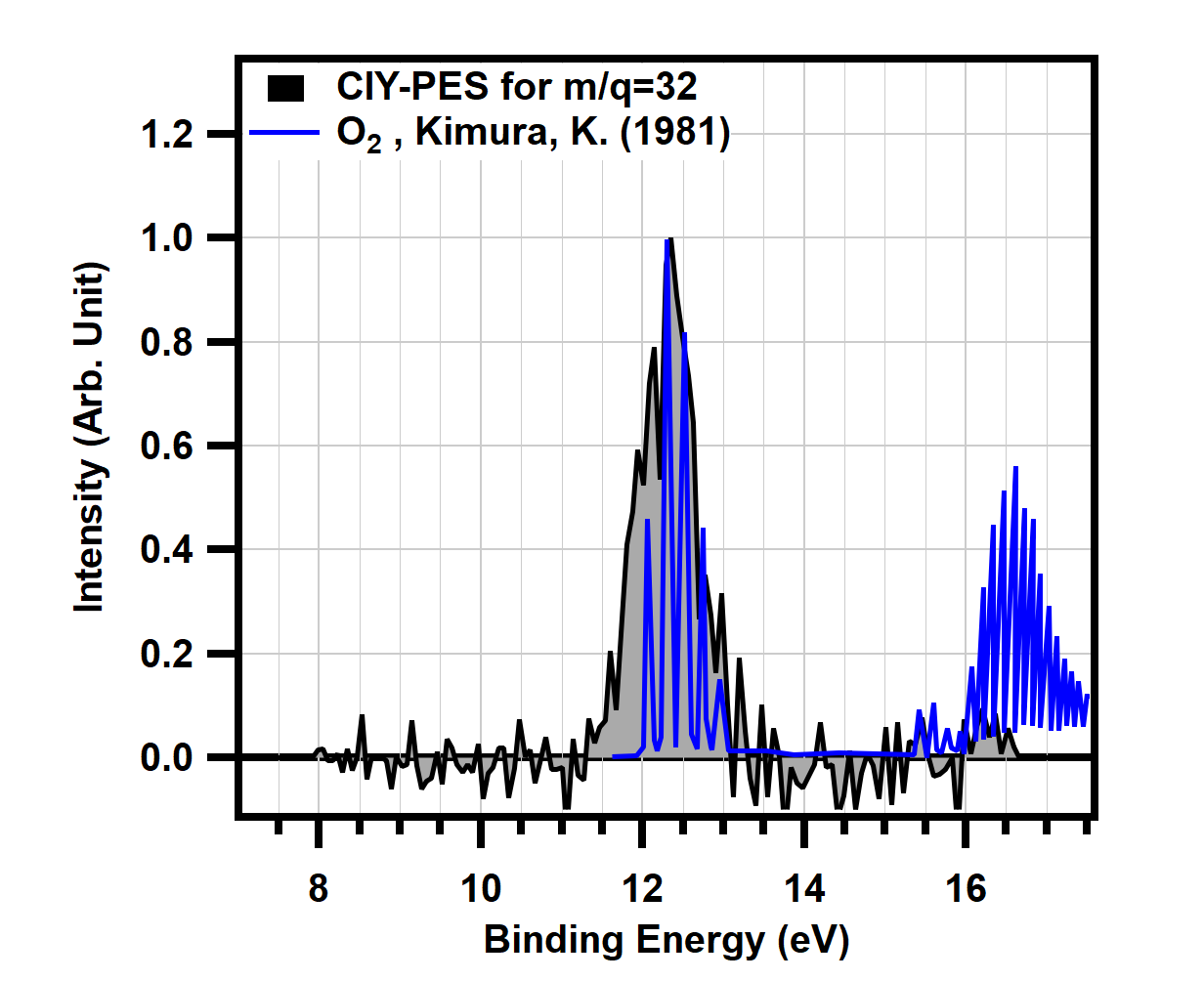}
\caption{The comparison of extracted coincidence ion yield PES at $40$ eV photon energy with reference spectra for ions with mass $32$ \cite{kimura1981}.}
\label{figS:O2}
\end{figure}

\begin{figure}
\centering
\includegraphics[width=0.66\textwidth]{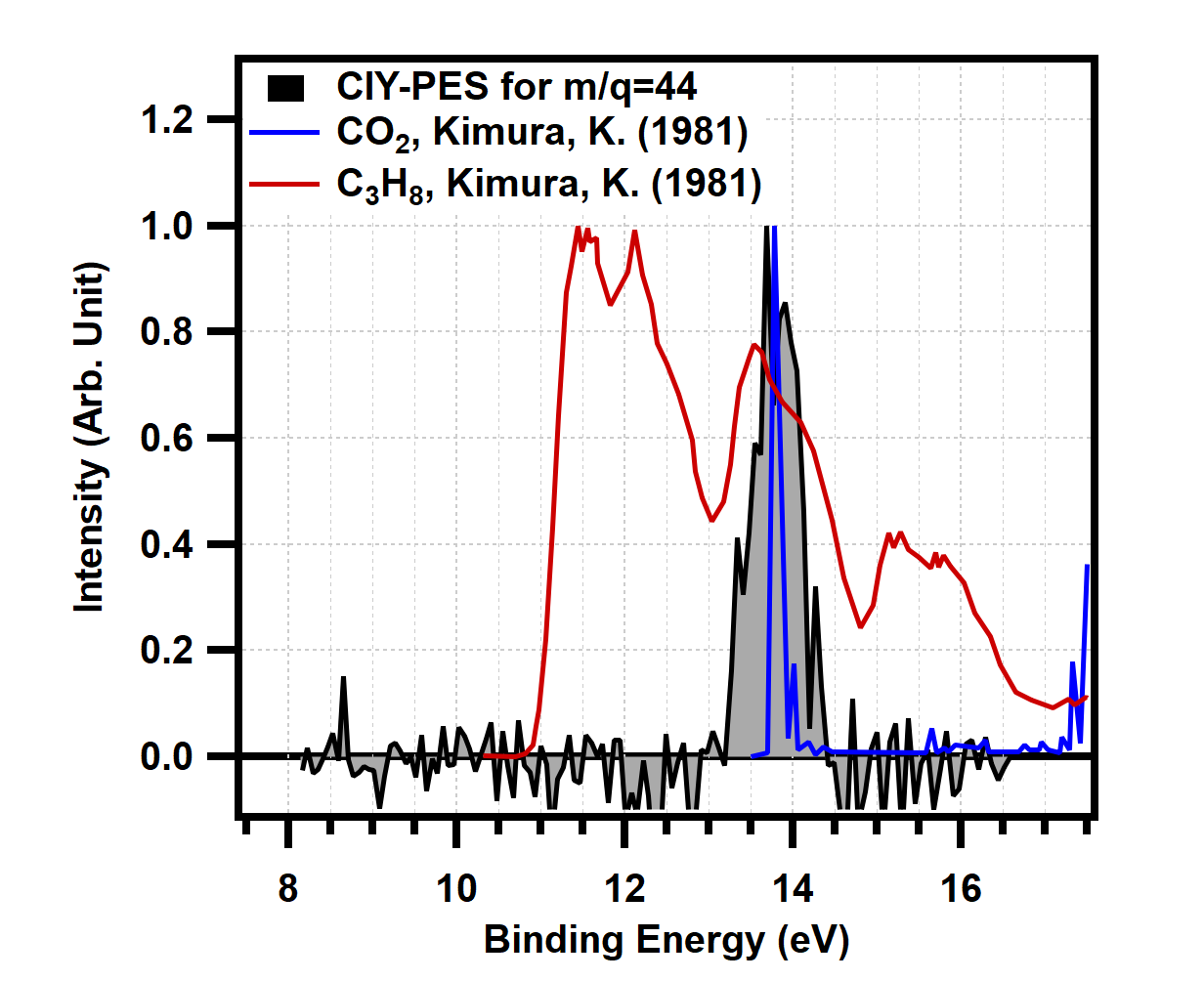}
\caption{The comparison of extracted coincidence ion yield PES at $40$ eV photon energy with reference spectra for ions with mass $44$ \cite{kimura1981}.} 
\label{figS:CO2}
\end{figure}

\begin{figure}
\centering
\includegraphics[width=0.64\textwidth]{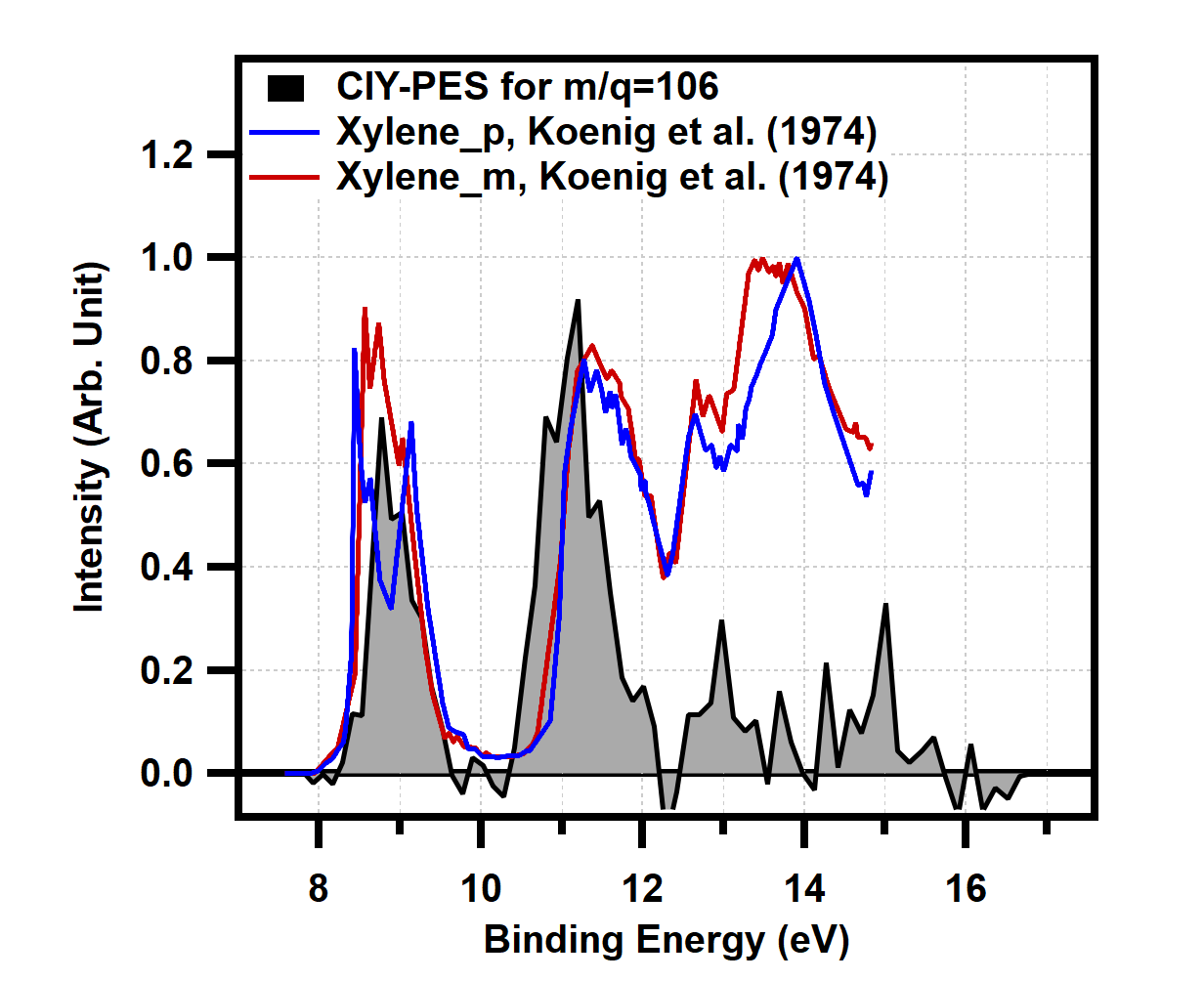}
\caption{The comparison of extracted coincidence ion yield PES at $40$ eV photon energy with reference spectra for ions with mass $106$ \cite{koenig1974}.}
\label{figS:C8H10_1}
\end{figure}

\pagebreak

\end{document}